\documentclass[aps,pre,onecolumn,letterpaper]{revtex4}
\usepackage{graphicx, bm, bbm,amsmath, amssymb,amsfonts, epsfig,color,comment}
\usepackage{float}
\usepackage[position=bottom, font=small]{subcaption}

\pdfoutput=1

\newcommand{\bq}{\begin{align}}
\newcommand{\eq}{\end{align}}

\begin{document}

\title{Enhancing wall-to-wall heat transport with unsteady flow perturbations}
\author{Silas Alben$^*$, Xiaojia Wang, and Nicole Vuong}
\affiliation{Department of Mathematics, University of Michigan,
Ann Arbor, MI 48109, USA}
\email{alben@umich.edu}

\begin{abstract}
We determine unsteady flow perturbations that are optimal for enhancing the rate of heat transfer between hot and cold walls  (i.e. the Nusselt number Nu), under the constraint of fixed flow power (Pe$^2$, where Pe is the P\'{e}clet number). The unsteady flows are perturbations of previously computed optimal steady flows and are given by eigenmodes of the Hessian matrix of Nu, the matrix of second derivatives with respect to amplitudes of flow mode coefficients. Positive eigenvalues of the Hessian correspond to increases in Nu by unsteady flows, and occur at Pe $\geq 10^{3.5}$ and within a band of flow periods $\tau \sim$ Pe$^{-1}$. For $\tau$Pe $\leq 10^{0.5}$, the optimal flows are chains of vortices that move along the walls or along eddies enclosed by flow branches near the walls. At larger $\tau$Pe the vorticity distributions are often more complex and extend farther from the walls. The heat flux is enhanced at locations on the walls near the unsteady vorticity. We construct an iterative time-spectral solver for the unsteady temperature field and find increases in Nu of up to 7\% at moderate-to-large perturbation amplitudes.  
\end{abstract}


\maketitle

\section{Introduction}

One of the most widely studied applications of fluid dynamics is convective heat transfer. This field is closely connected to fundamental areas of fluid dynamics such as boundary layers, turbulence, and multiphase flows \cite{layton1988history,kaviany1994principles}. Many studies have considered how to manipulate boundary layers and turbulence to enhance convective heat transfer from solid surfaces \cite{webb2005enhanced}. ``Passive" techniques include roughening the solid surfaces and inserting vortex generators or other obstacles to enhance fluid and thermal mixing within the boundary layer \cite{webb2005enhanced,glezer2016enhanced}. ``Active" techniques involve applying local forcing to the fluid (or the boundaries) using external electric fields, jet impingement, fluid injection, stirring, or surface vibrations \cite{yabe1996active,laohalertdecha2007review,fang2013active,leal2013overview,shank2023review,webb2005enhanced,wang2025vortices}.

Hassanzadeh et al. \cite{hassanzadeh2014wall} formulated an optimization problem in order to study upper bounds on the rate of convective heat transfer in a simple geometry. They considered two-dimensional steady incompressible fluid flows in a layer between hot and cold horizontal walls. They solved the Euler-Lagrange equations numerically, and in certain limits analytically, to determine the flows that maximize the rate of heat transfer (the Nusselt number Nu) subject to a given rate of viscous power dissipation (Pe$^2$, with Pe the P\'{e}clet number). The problem setup was motivated by the longstanding search for upper bounds on heat transfer in natural (Rayleigh-B\'{e}nard) convection \cite{sondak2015optimal,wen2020steady,wen2022steady,lohse2024ultimate}. The optimal flows in \cite{hassanzadeh2014wall} are only constrained by incompressibility and fixed power dissipation, not by the buoyancy-driven Navier-Stokes equations (the Oberbeck-Boussinesq equations) of natural convection. In fact, the optimal flows {\it are} solutions to the Navier-Stokes equations with a suitable forcing term that may be taken as a model for the forcing in active heat transfer enhancement. More generally, the structures of these optimal flows improve our understanding of the fluid mechanics of efficient heat transfer in both natural and forced convection.

The ``wall-to-wall" optimization problem of \cite{hassanzadeh2014wall} was studied and extended by several subsequent works. Souza et al. extended the work to no-slip instead of free-slip boundary conditions, and higher power dissipation (i.e. Pe) \cite{souza2016optimal,souza2020wall}. They also studied optimal transport in the Lorenz equations, a low-mode truncation of the wall-to-wall problem \cite{souza2015maximal,souza2015transport}. Tobasco and Doering constructed analytical branching flows that saturate an upper bound on the rate of heat transfer up to a logarithmic correction \cite{tobasco2017optimal,doering2019optimal}. Alben computed 2D optimal flows using an adjoint method up to Pe = 10$^{7.5}$ and found branched structures for 
Pe $\gtrsim$ 10$^{4.5}$ \cite{alben2023transition}.
Previously, Motoki et al. had solved the Euler-Lagrange equations for optimal steady 3D flows computationally using a continuation method \cite{motoki2018optimal, motoki2018maximal}. For the wall-to-wall problem they found a bifurcation from 2D convection rolls to 3D optima at Pe $\approx$ 80 \cite{motoki2018maximal}. The flows show a tree-like pattern with branching near the walls for Pe $\gtrsim 500$, much lower than in 2D. Kumar constructed 3D branching ``pipe" flows analytically that attain the upper bound without the logarithmic correction \cite{kumar2024three}. The upper bound was also found for 2D walls with elongated convection rolls by allowing for wavy wall shapes \cite{alben2024optimal}. The same optimization framework was used to study optimal flows through channels \cite{alben2017improved} and other geometries \cite{alben_2017}. A body of work has also found improved heat transfer in natural convection from rough walls \cite{toppaladoddi2017roughness,toppaladoddi2021thermal}. Other recent work has considered optimal cooling of the interior of domains \cite{marcotte2018optimal,song2023bounds}.
 
In this work we extend the wall-to-wall optimization problem to the unsteady case, in 2D space. Solving the discretized steady advection-diffusion equation in the advection-dominated regime requires a fine mesh in order to resolve very thin temperature boundary layers and thin plumes that move far from the boundary, into the main flow. In the adjoint-based optimization method of \cite{alben2023transition}, the problem was formulated as an unconstrained optimization problem in which the Nusselt number was maximized using BFGS, a Hessian-free iterative method. At each iteration, a backtracking line search was performed in which the steady advection-diffusion equation was solved multiple times, to find a flow that yields a larger Nu. O($10^3$--$10^4$) iterates were computed for each initial guess for the optimal flow. Multiple initial guesses were used at each Pe to search for multiple local optima. A wide range of Pe values was used, [0, 10$^{7.5}$]. Altogether there were many solves of the advection-diffusion equation and the total computational cost was substantial.

In the unsteady case, this optimization approach would be orders of magnitude more expensive. Adding the time dimension to the two spatial dimensions makes solving the advection-diffusion equation much more expensive. An explicit time-stepping scheme converges very slowly to the time-periodic solution in the advection-dominated regime. Instead, we adopt an implicit time-spectral method here, which results in a sparse linear system with O(10$^6$--10$^7$) unknowns, and requires a iterative solver. The condition number of the linear system is large, again because we work in the advection-dominated regime, so many iterations are required in some cases. In addition, the search occurs in a much higher dimensional space than the steady case, with the additional coefficients needed to describe the time dependence of the flow modes.
In order to mitigate the computational cost, we restrict the optimization problem to the limit of small unsteady flow perturbations, using the approach of \cite{alben2024enhancing}. However, we evaluate the performance of the optimal perturbations with amplitudes ranging from small to large.

The unsteady 2D wall-to-wall heat transfer optimization problem was considered previously by Souza \cite{souza2016optimal}. A gradient ascent method was proposed but results were limited to the steady case due to computational challenges. For the unsteady problem, a low-mode approximation was considered: the Lorenz equations, with one flow mode and two temperature modes. The ``background" method and an optimal control method were used to show that steady flows are optimal in this approximation. The same methods were used to show that steady flows are also optimal for the ``double Lorenz" equations, with four flow modes and four temperature modes. In this paper, we investigate whether steady flows are optimal for the unsteady advection-diffusion equation.

We compute optimal flow modes in the limit of small unsteady perturbations of the currently best known steady 2D flows from \cite{alben2023transition}.
We show that unsteady flows outperform the steady flows above a critical Pe value. The unsteady flows are defined by a finite set of eigenmodes of a Hessian matrix. We test the unsteady perturbations with amplitudes ranging from small to large and find modest but noticeable increases in the rate of heat transfer (Nu). 

In section \ref{sec:Setup} we introduce the problem. Section \ref{sec:pert} explains the perturbative approach for computing optima. The numerical discretization is given in section \ref{sec:Disc}, followed by the modal expansions of the perturbations in section \ref{sec:Modes}. Section
\ref{sec:eval} shows the perturbations that can increase Nu, indicated by positive eigenvalues of the Hessian in the quadratic term in the expansion of Nu. The corresponding eigenmodes' vorticity and temperature fields are described. Section \ref{sec:UnsteadySims} shows the temperature fields and Nu values when the perturbations are employed with a range of amplitudes from small to large. Section \ref{sec:Conclusions} gives the conclusions. 

\section{Problem setup \label{sec:Setup}}

\begin{figure}
    \centering
\includegraphics[width=0.9\textwidth]{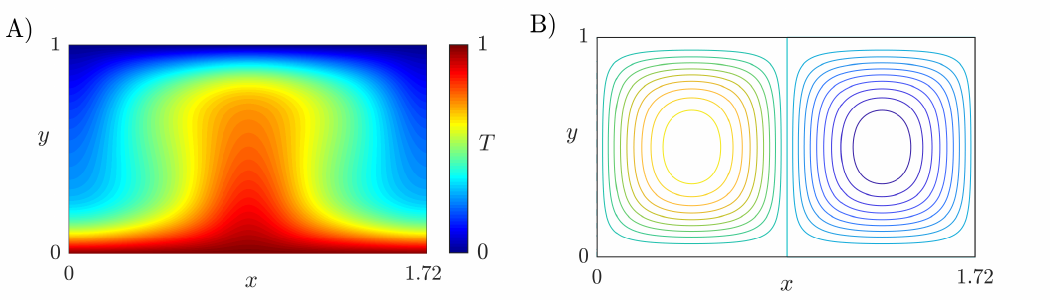}
    \caption{\footnotesize Temperature field (panel A) corresponding to a horizontally periodic flow with streamlines shown in panel B. The flow has two counter-rotating convection rolls per unit cell and Pe = $10^2$.}
    \label{fig:ExampleTempFlowFig}
             \vspace{-.10in}
\end{figure}

The ``wall-to-wall" problem has been described in a series of works \cite{hassanzadeh2014wall,souza2016optimal,tobasco2017optimal,motoki2018maximal,doering2019optimal,souza2020wall,alben2023transition} so we give a brief description of the problem---its geometry, equations, and boundary conditions.
A layer of fluid is contained between horizontal walls at $y = 0$ and 1, with temperatures 1 and 0 respectively (an example is shown in figure \ref{fig:ExampleTempFlowFig}A). We have chosen to nondimensionalize all quantities with dimensions of length (including $y$) by the layer height $H$ and quantities with units of temperature by the temperature difference between the walls $\Delta T$.
Given an incompressible flow field $\mathbf{u} = (u(x,y,t),v(x,y,t))$, we solve the unsteady advection-diffusion equation for the temperature $T$:
\begin{align}
    \partial_tT + u\partial_x T + v\partial_y T - \nabla^2T = 0. \label{AdvDiff}
\end{align}
In the dimensional version of (\ref{AdvDiff}), the prefactor of $\nabla^2T$ is the thermal diffusivity $\kappa$; it is set to unity in the dimensionless equation (\ref{AdvDiff}), corresponding to nondimensionalizing velocities by $\kappa/H$. Figure \ref{fig:ExampleTempFlowFig}A
shows the temperature field for an example flow in panel B that is steady and horizontally periodic with period 1.72. The flow has counter-rotating convection rolls (counterclockwise on the left, clockwise on the right) that result in an upward plume of hot fluid in the middle of the $x$ range in panel A and a downward cold plume that is centered at the periodic boundary at $x$ = 0 and $x$ = 1.72.

We search for a 2D incompressible flow field that maximizes the Nusselt number Nu, the rate of heat transfer out of the bottom wall, averaged over time and horizontal period length:
\begin{align}
    \mbox{Nu} &= \frac{1}{\tau}\int_0^\tau \frac{1}{L_x}\int_0^{L_x} -\partial_y T \Big|_{y = 0} dx \, dt. \label{Nu}
\end{align}
Here we assume a time-periodic temperature field with period $\tau$ (i.e. due to a flow with the same period).
Using (\ref{AdvDiff}) and the divergence theorem one can show that Nu is also the average rate of heat transfer into the top wall. We maximize Nu over flows that satisfy certain constraints. First, we require the flow to be incompressible, automatically imposed by setting $(u,v) = \mathbf{u} = (\partial_y\psi,-\partial_x\psi)$, for a given stream function $\psi(x,y,t)$. We require zero flow velocity (the no-slip condition) at the walls: $\psi = \partial_y\psi = 0$ at $y = 0$ and $\psi = \psi_{top}$ and $\partial_y\psi = 0$ at $y = 1$. The constant $\psi_{top}$ is the net horizontal fluid flux through the channel. We will take $\psi_{top} = 0$ to be consistent with the previous steady optimal flows, discussed further below. One of the main constraints is a fixed average rate of viscous energy dissipation, i.e. the time- and space-averaged power needed to maintain the flow, in dimensionless form,
\begin{align}
   \mbox{Power} = \frac{1}{\tau} \int_0^\tau \int_0^1 \frac{1}{L_x} \int_0^{L_x}  2e_{ij}^2\, dx\, dy\, dt, \label{PowerInit}
\end{align}
with the spatial integral taken over a period cell. Here $e_{ij} = (\partial_{x_i}u_j + \partial_{x_j}u_i)/2$ is the symmetric part of the velocity gradient tensor and $e_{ij}^2$ is the sum of the squares of its entries. Thus
\begin{align}
    2 e_{ij}^2 &= 2 \partial_x u^2 + (\partial_y u + \partial_x v)^2 + 2\partial_y v^2.  
\end{align}
We have nondimensionalized the space- and time-averaged power in (\ref{PowerInit}) by $\mu \kappa^2/H^4$ where $\mu$ is the fluid viscosity.  
We now write $u$ and $v$ in terms of $\psi$ to obtain:
\begin{align}
   \mbox{Power} &= \frac{1}{\tau}\int_0^\tau \int_0^1 \frac{1}{L_x}\int_0^{L_x} \left(\partial_{xx}\psi - \partial_{yy}\psi\right)^2 + 4\partial_{xy}\psi^2  \,dx\,dy\,dt = \frac{1}{\tau}\int_0^\tau \int_0^1 \frac{1}{L_x} \int_0^{L_x} \left(\nabla^2\psi\right)^2  \,dx\,dy\,dt. \label{Power}
\end{align}
The second equality in (\ref{Power}) holds when the flow is periodic or has zero tangential component at the boundary, as is the case here \cite[Art. 329]{lamb1932hydrodynamics}. 
The basic problem is to maximize Nu over a set of incompressible flows with power = Pe$^2$, a constant:
\begin{align}
\max_\psi \mbox{Nu}(\psi) \quad
\mbox{subject to} \quad \begin{cases}
    &\partial_tT + \partial_y\psi\partial_x T -\partial_x\psi\partial_y T - \nabla^2T = 0, \\
&\frac{1}{\tau}\int_0^\tau \int_0^1 \frac{1}{L_x} \int_0^{L_x} \left(\nabla^2\psi\right)^2  \,dx\,dy\,dt = \mbox{Pe}^2 \end{cases} \label{ContOpt}
\end{align}
where Pe stands for ``P\'{e}clet number," a measure of the dimensionless flow velocity. The boundary conditions for $T$ in problem (\ref{ContOpt}) are given above equation (\ref{AdvDiff}). The class of $\psi$ is left unspecified because different choices are possible. In \cite{alben2023transition} we computed optimal {\it steady} flows $\psi$, with no-slip boundary conditions at the top and bottom walls, and horizontally periodic.

In this paper, we compute optimal {\it unsteady} flows, assuming they are small unsteady perturbations of the optimal steady flows. Extending the algorithm in \cite{alben2023transition} directly to unsteady flows is too expensive, as we will discuss later, but the perturbative assumption makes the computations feasible. However, we will evaluate the resulting unsteady flow perturbations at both small and large amplitudes. The optimal steady flows in \cite{alben2023transition} were computed for Pe $\in [0, 10^{7.5}]$, and the horizontal period $L_x$ was optimized together with $\psi$. $\psi_{top} = 0$ was found to be optimal in all cases. Since our unsteady flows will be perturbations of the steady optima, for simplicity we will assume $\psi_{top} = 0$ and the flow is periodic in $x$ with the same period as the optimal steady flow at a given Pe, $L_x$ given in \cite{alben2023transition}.
 
In summary, the problem has the geometry and boundary conditions of Rayleigh-B\'{e}nard convection, but instead of solving the Oberbeck-Boussinesq equations for the flow field, we
determine the flow field that maximizes the Nusselt number, the average rate of heat transfer from the boundaries. The optimal flows are solutions to the 
Navier-Stokes equations with a certain force per unit volume distributed over the flow domain, in other words, to a forced convection flow. The forcing function is not computed explicitly here, but a method for doing so is discussed in appendix A of \cite{alben2017improved}. However, the average rate of work done by the forcing function corresponds to the average rate of viscous dissipation, Pe$^2$.

\section{Unsteady perturbations of steady optima \label{sec:pert}}

The expansion of the unsteady flows as perturbations of the steady flows, the corresponding Nu expansion, and the computation of optimal unsteady flow modes follows the same procedure as in \cite{alben2024enhancing} with a few adjustments. The steady flows in \cite{alben2024enhancing} were different---horizontal flows through a channel with cold fluid flowing inward at the left boundary, heated walls (with $T$ = 1) at the top and bottom boundaries, and outflow at the right boundary, instead of the horizontally periodic problem here in which heat moves from the bottom to the top. 

\begin{figure}
    \centering
    \includegraphics[width=1\textwidth]{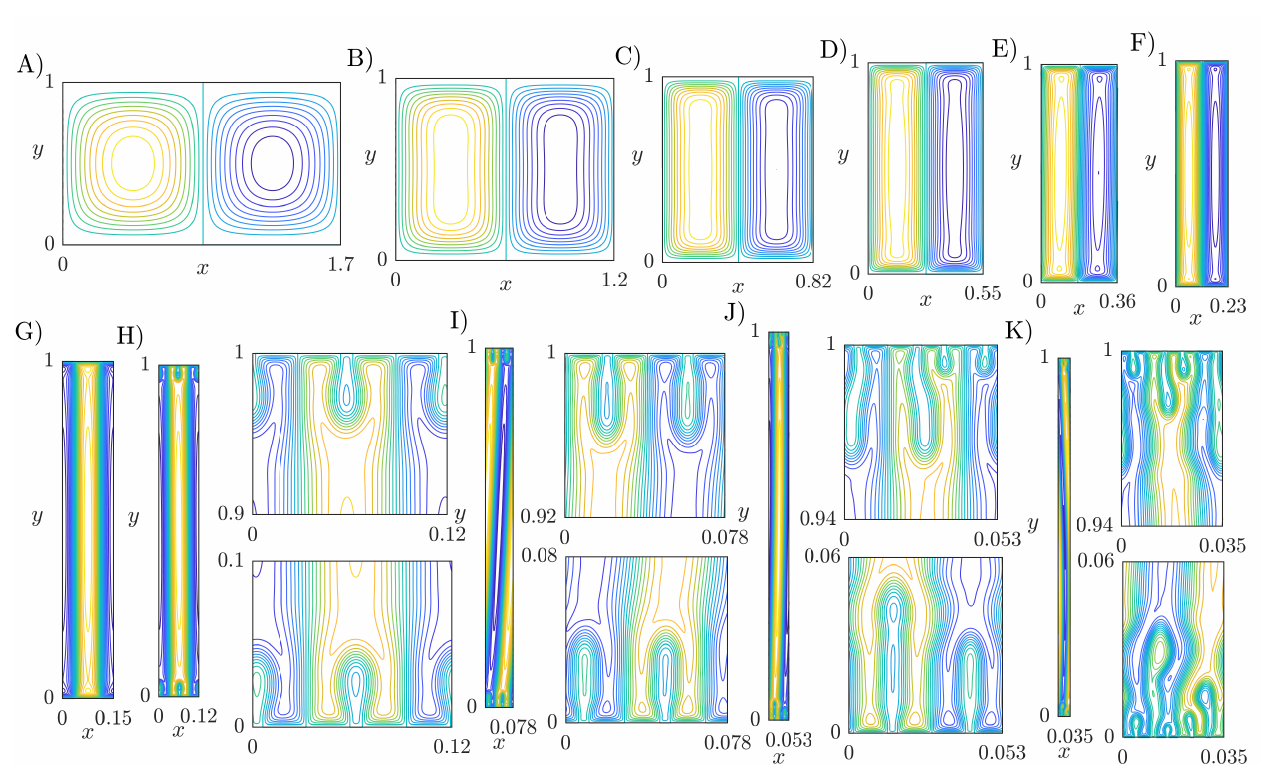}
    \caption{\footnotesize Streamlines of optimal steady flows computed in \cite{alben2023transition} at Pe = 10$^2$ (panel A), 10$^{2.5}$ (B), 10$^3$ (C), 10$^{3.5}$ (D),
    10$^4$ (E), 10$^{4.5}$ (F), 10$^5$ (G), 10$^{5.5}$ (H), 10$^6$ (I), 10$^{6.5}$ (J), and 10$^7$ (K). For panels H--K, inset panels to the right of the main flows show the streamlines near the top and bottom walls.}
    \label{fig:StaticFlowsFig}
             \vspace{-.10in}
\end{figure}

In figure \ref{fig:StaticFlowsFig} we show the base states of the perturbation expansion---the optimal steady flows from \cite{alben2023transition} at Pe ranging from 10$^2$ to 10$^7$. These are the best flows found from 30--50 random initializations using a BFGS iterative method. The flows consist of rectangular convection rolls with horizontal period $L_x \sim$ Pe$^{-0.36}$. Small closed eddies occur near the walls at Pe = $10^4$ (panel E) and above, and branching flows are optimal at Pe $> 10^5$ (panels H--K). Up to and including Pe = $10^4$ (A--E), most initializations converged to the global optimum, with just a few resulting in a suboptimal flow that is slightly perturbed, usually near the walls. Between Pe = 10$^{4.5}$ and 10$^6$ (F--I), the global optimum was found multiple times, but with decreasing frequency as Pe increased. Here a wider range of suboptimal flows was found, with different branching patterns near the walls. At 
Pe = 10$^{6.5}$ and 10$^7$ (J and K), the branching patterns are more complex with multiple levels, and several flows with small to moderate variations in branching are within 0.1\% of the maximum Nu found. Only the flows with the maximum Nu are shown in figure \ref{fig:StaticFlowsFig}, and these are used as the base flows for the perturbation expansions.

\begin{figure}
    \centering
    \includegraphics[width=1\textwidth]{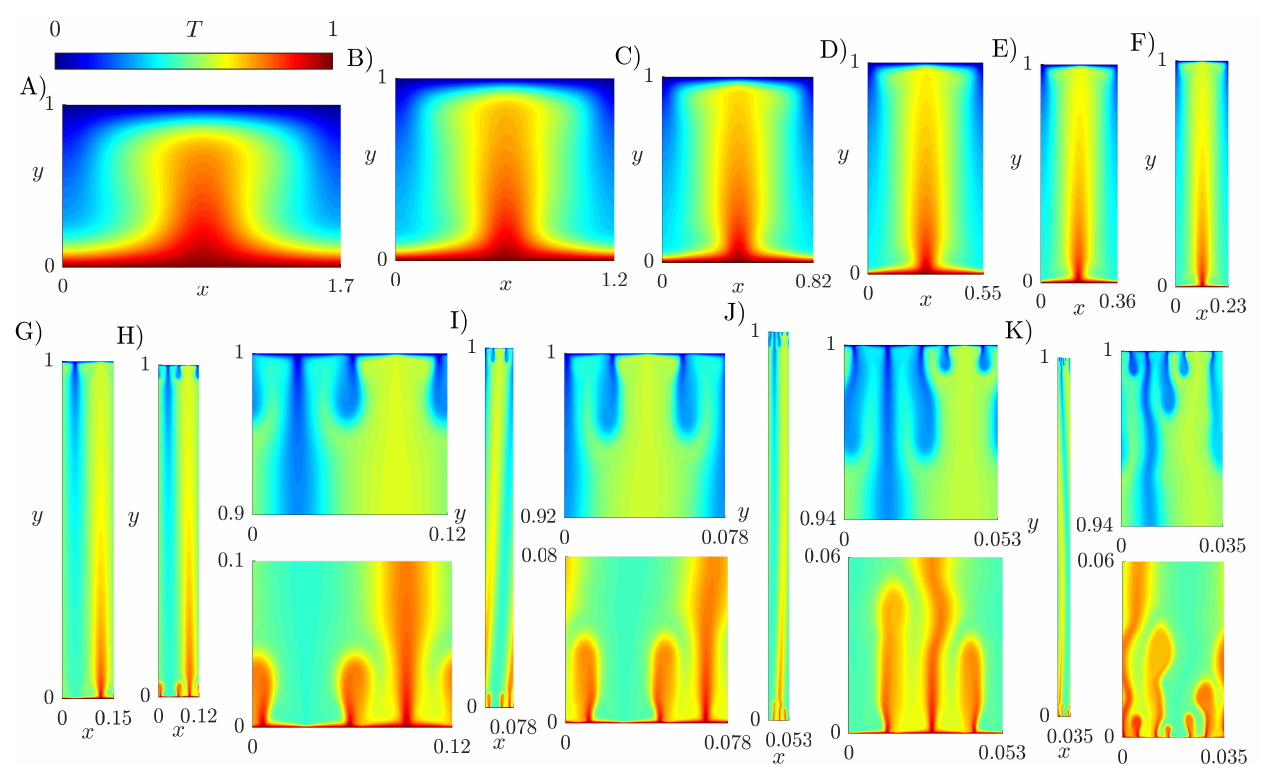}
    \caption{\footnotesize Temperature fields corresponding to the flows in figure \ref{fig:StaticFlowsFig}.}
    \label{fig:StaticTempFig}
             \vspace{-.10in}
\end{figure}

Figure \ref{fig:StaticTempFig} shows the temperature fields corresponding to the optimal steady flows. Upward and downward plumes are located at the streamlines that emanate from the walls, separating neighboring convection rolls or enclosed eddies. Computing Nu accurately requires resolving very thin temperature boundary layers with fine meshes as discussed below in section \ref{sec:Disc}.  

\begin{figure}
    \centering
    \includegraphics[width=0.7\textwidth]{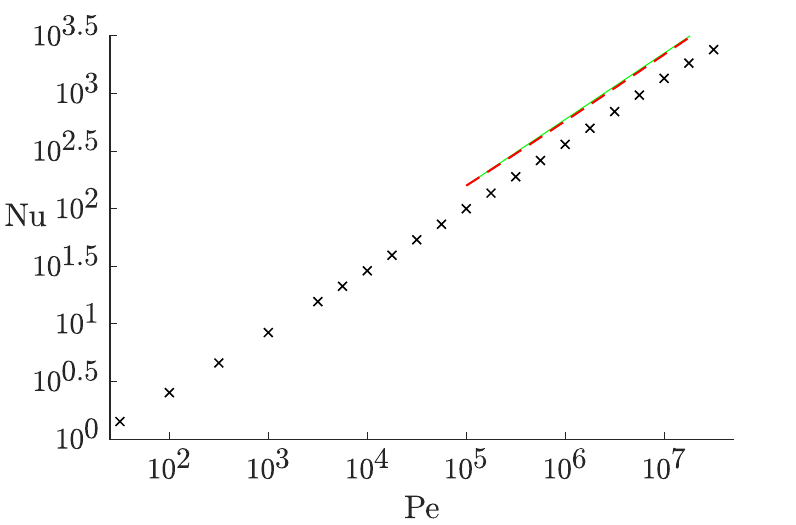}
    \caption{\footnotesize Nusselt number versus P\'{e}clet number (black crosses) for steady optimal flows and temperature fields including those shown in figures \ref{fig:StaticFlowsFig} and \ref{fig:StaticTempFig}. The green and red lines show Nu $\sim$ Pe$^{0.575}$ and Nu $\sim$ Pe$^{2/3}$(log Pe)$^{-4/3}$ respectively. }
    \label{fig:NuStVsPeFig}
             \vspace{-.10in}
\end{figure}

Figure \ref{fig:NuStVsPeFig} shows the values of Nu for the optimal flows over a range of Pe, including those in figure \ref{fig:StaticFlowsFig}. The green and red lines show Nu scalings proportional to Pe$^{0.575}$ and Pe$^{2/3}$(log Pe)$^{-4/3}$ respectively. The latter form was given for analytically-constructed flows with a self-similar branched structure in \cite{tobasco2017optimal,doering2019optimal}. One of the main questions we address is whether we can find unsteady flows with larger Nu than those of the steady optima in figure \ref{fig:NuStVsPeFig}.

At a given Pe, we denote the optimal steady flow as $\psi_0(x,y)$ and consider an unsteady flow of the form
\begin{align}
\psi(x,y,t) = \frac{1}{\sqrt{1+c_1\epsilon^2}}\psi_0(x,y) + \frac{\epsilon}{\sqrt{1+c_1\epsilon^2}}f_A(x,y)\cos(2\pi t/\tau)
+ \frac{\epsilon}{\sqrt{1+c_1\epsilon^2}}f_B(x,y)\sin(2\pi t/\tau). \label{psipert}
\end{align}
Here $0 < \epsilon \ll 1$ and the 
constant $c_1$ will be chosen so that $\psi$ has the same time-averaged power as $\psi_0(x,y)$, Pe$^2$. 
We will now compute Nu for $\psi$ by expanding both $\psi$ and $T$ in powers of $\epsilon$,
\begin{align}
    \psi(x,y,t) &= \psi_0(x,y) + \epsilon\psi_1(x,y,t) + \epsilon^2\psi_2(x,y,t) + \ldots \label{Psiexpn}\\
    T(x,y,t) &= T_0(x,y) + \epsilon T_1(x,y,t) + \epsilon^2 (T_{2s}(x,y) + T_{2u} (x,y,t)) + \ldots, \label{TexpnEps}
\end{align}
plugging both series into equation (\ref{AdvDiff}), and solving the equation at each power of $\epsilon$. Then we plug the $T$ expansion
(\ref{TexpnEps}) into (\ref{Nu}) to obtain a similar expansion for Nu:
\begin{align}
    \mbox{Nu} &= \mbox{Nu}_0 + \epsilon \mbox{Nu}_1 + \epsilon^2 \mbox{Nu}_2 + \ldots. \label{Nuexpn}
\end{align}
In (\ref{TexpnEps}) we have separated $T_2$ into its time average $T_{2s}$ (the steady part) and the remainder $T_{2u}$ (the unsteady part); only the steady part will be needed for Nu, which is a time average.
At $O(\epsilon^0)$ equation (\ref{AdvDiff}) is
\begin{align}
    \partial_y\psi_0 \partial_xT_0 - \partial_x\psi_0\partial_yT_0 - \nabla^2T_0 = 0 \label{T0}
\end{align}
and thus Nu$_0$ is the Nusselt number for the optimal steady flow $\psi_0$.
At $O(\epsilon^1)$ equation (\ref{AdvDiff}) is
\begin{align}
    \partial_t T_1 + \partial_y\psi_0 \partial_xT_1 - \partial_x\psi_0\partial_yT_1 - \nabla^2T_1 = - \partial_y\psi_1 \partial_xT_0 + \partial_x\psi_1\partial_yT_0. \label{T1}
\end{align}
Since the right-hand side has components proportional to $\cos(2\pi t/\tau)$ and $\sin(2\pi t/\tau)$, the solution can be written as $T_1 = T_{1A}(x,y)\cos(2\pi t/\tau) + T_{1B}(x,y)\sin(2\pi t/\tau)$. Inserting this expression
into (\ref{T1}) gives a coupled system for $T_{1A}$ and $T_{1B}$:
\begin{align}
    \frac{2\pi}{\tau}T_{1B} + \partial_y\psi_0 \partial_xT_{1A} - \partial_x\psi_0\partial_yT_{1A} - \nabla^2T_{1A} &= -\partial_y f_A \partial_xT_0 +  \partial_x f_A\partial_yT_0. \label{T1A} \\
     -\frac{2\pi}{\tau} T_{1A} + \partial_y\psi_0 \partial_xT_{1B} - \partial_x\psi_0\partial_yT_{1B} - \nabla^2T_{1B} &= -\partial_y f_B \partial_xT_0 + \partial_x f_B\partial_yT_0.
    \label{T1B}
\end{align}

Since $T_1$ has zero time-average, Nu$_1 = 0$. We need to compute higher-order terms in Nu to determine whether the unsteady flow perturbation can increase Nu. 
At $O(\epsilon^2)$ equation (\ref{AdvDiff}) is
\begin{align}
    \partial_t T_2 + \partial_y\psi_0 \partial_xT_2 - \partial_x\psi_0\partial_yT_2 - \nabla^2T_2 = - \partial_y\psi_1 \partial_xT_1 + \partial_x\psi_1\partial_yT_1 - \partial_y\psi_2 \partial_xT_0 + \partial_x\psi_2\partial_yT_0. \label{T2}
\end{align}
Equation (\ref{psipert}) gives $\psi_2 = -c_1\psi_0/2$. Since $\psi_1$ and $T_1$ are proportional to $\cos(2\pi t/\tau)$ and $\sin(2\pi t/\tau)$, the right side of (\ref{T2}) has steady terms and terms proportional to 
$\cos(4\pi t/\tau)$ and $\sin(4\pi t/\tau)$.
The unsteady terms have zero time-average, so only the steady terms make a nonzero contribution to Nu$_2$. We compute it by first solving for $T_{2s}$:
\begin{align}
    \partial_y\psi_0 \partial_xT_{2s} - \partial_x\psi_0\partial_yT_{2s} - \nabla^2T_{2s} = - \frac{1}{2}\partial_yf_A \partial_xT_{1A} + \frac{1}{2}\partial_xf_A\partial_yT_{1A}& \nonumber\\
    - \frac{1}{2}\partial_yf_B \partial_xT_{1B} + \frac{1}{2}\partial_xf_B\partial_yT_{1B} + \frac{c_1}{2}\partial_y\psi_0 \partial_xT_0 - \frac{c_1}{2}\partial_x\psi_0\partial_yT_0&, \label{T2s}
\end{align}
and then computing
\begin{align}
    \mbox{Nu}_2 &= \frac{1}{L_x}\int_0^{L_x} -\partial_y T_{2s} \Big|_{y = 0}. \label{Nu2}
\end{align}
The time average is dropped in (\ref{Nu2}) because the integrand is steady.
To summarize, the leading-order change in Nu due to the unsteady flow perturbation is $\epsilon^2$Nu$_2$, and to calculate Nu$_2$ we need to calculate $T_{1A}$, $T_{1B}$, and $T_{2s}$. This requires solving three elliptic PDEs---(\ref{T1A}), (\ref{T1B}), and (\ref{T2s})---two of which are coupled. In the next section (\ref{sec:Disc}) we discuss the numerical discretization of the equations and in the following section (\ref{sec:Modes}) we explain how the calculation of 
Nu$_2$ is used to determine the optimal unsteady perturbation as a superposition of flow modes.

\section{Discretization \label{sec:Disc}}
We solve (\ref{T1A}), (\ref{T1B}), and (\ref{T2s}) by representing each component of $\psi$ and $T$ in (\ref{Psiexpn}) and  (\ref{TexpnEps}) on tensor product grids in $(x,y)$ space that are concentrated at the boundaries, the same as in \cite{alben2023transition}. We start with unstretched coordinates $p$ and $Y$. We define uniform grids on $[0, 1]$ in $p$ and $Y$ with spacings $1/m$ and $1/n$ respectively, where $m$ = 256 and $n$ = 256, 512, or 1024. We then map these points to $(x,y)$ space:
\begin{align}
    x = L_x p \quad ; \quad y = Y - \eta\frac{1}{2\pi}\sin 2\pi Y
\end{align}
with $\eta$ a stretching factor that is set to 0.997. We obtain a stretched grid with points concentrated near the boundary in $y$, to resolve the boundary layers. The grid spacing in $y$ ranges from about $(1-\eta)/n \approx 3\times 10^{-6}$--$10^{-5}$ at the boundaries to about $(1+\eta)/n \approx 2\times 10^{-3}$--$8\times 10^{-3}$ at the domain center. We discretize all spatial derivatives with second-order finite differences using three or four grid points, one-sided at the boundaries.
Integrals such as those in (\ref{Nu}) and (\ref{Power}) are discretized with the trapezoidal rule.

\section{Flow modes \label{sec:Modes}}
Our overall goal is to maximize Nu over a large space of unsteady flow perturbations. The particular form of the perturbation in (\ref{psipert}) is determined by the functions $f_A$ and $f_B$ and the scalar $\tau$. To represent a wide range of flows, we expand $f_A$ and $f_B$ using a large number of products of Fourier modes in $x$ and Chebyshev polynomials in $y$, the same as for the steady optima in \cite{alben2023transition}. The $2M+1$ Fourier modes are arranged as alternating cosines and sines, from small to large wave numbers:
\begin{align}
X_{2j+1}(x) = \cos(2\pi j x/L_x) \;, j = 0,\ldots, M\quad ; \quad X_{2j}(x) = \sin(2\pi j x/L_x), j = 1,\ldots, M.
\end{align}
We take $M = 5m/64$, much smaller than $m$ so the modes are resolved well by the horizontal mesh. The vertical modes $\{Y_k(y), k = 1, \ldots, N-3$\}, are linear combinations of Chebyshev polynomials that satisfy no-slip at the walls. They are generated using an orthogonal projection procedure described in \cite[App. B]{alben2023transition}. We take $N = n/8$, so the most oscillatory mode ($k = N-3$) is resolved well by the vertical mesh.

We form all the products $X_j(x)Y_k(y)$ on the spatial grid and assemble them as $(2M+1)(N-3)$ columns of a matrix $\mathbf{Z}$. The problem is further simplified by making the flow modes orthogonal with respect to the power inner product (\ref{Power}), in discretized form.  For a discretized flow given as a column vector  $\mathbf{Z}_i$ multiplied by either $\cos (2\pi t/\tau)$ or  $\sin (2\pi t/\tau)$, we can write the discretized power integral (\ref{Power}) as a weighted inner product $(\sqrt{\mathbf{W}}\mathbf{L}\mathbf{Z}_i)^T\sqrt{\mathbf{W}}\mathbf{L}\mathbf{Z}_i$, where $\mathbf{L}$ is the discrete Laplacian and $\mathbf{W}$ is the diagonal matrix of weights corresponding to the compound trapezoidal rule for the spatial integrals in (\ref{Power}), multiplied by 1/2 for the time integral.
We perform a QR factorization
\begin{align}
\frac{1}{\mbox{Pe}}\mathbf{\sqrt{W}L}\mathbf{Z} = \mathbf{Q}\hat{\mathbf{R}}.
\end{align}
Then $\mathbf{V} \equiv \mathbf{Z}\hat{\mathbf{R}}^{-1}$ has columns $\mathbf{V}_j$ that obey
$(\sqrt{\mathbf{W}}\mathbf{L}\mathbf{V}_i)^T\sqrt{\mathbf{W}}\mathbf{L}\mathbf{V}_j$ = Pe$^2\delta_{ij}$. Thus the $\mathbf{V}_j$ are orthogonal with respect to the power inner product and have power Pe$^2$. These are the modes that are used as a basis for $f_A$ and $f_B$ in the unsteady part of the flow given by (\ref{psipert}). 

We use the modes of $\mathbf{V}$ to write a discrete version of the flow in (\ref{psipert}):
\begin{align}
\mathbf{\Psi} = \frac{1}{\sqrt{1+\|\mathbf{a}\|^2}} \mathbf{\Psi}_0 + \sum_{j = 1}^{N_m} \frac{\mathrm{a}_{j}}{\sqrt{1+\|\mathbf{a}\|^2}}
\mathbf{V}_j\cos\left(\frac{2\pi t}{\tau}\right) + 
\sum_{j = 1}^{N_m} \frac{\mathrm{a}_{j+N_m}}{\sqrt{1+\|\mathbf{a}\|^2}}
\mathbf{V}_j\sin\left(\frac{2\pi t}{\tau}\right).\label{Psi}
\end{align}
Here $\mathbf{V}_j$ are columns of $\mathbf{V}$ representing $N_m$ different spatial modes and $\mathbf{a}$ is a vector of coefficients for the modes. \textcolor{black}{The terms in (\ref{Psi}) are mutually orthogonal with respect to the power inner product.} Those with different temporal functions are orthogonal due to the time integral in (\ref{Power}), and those with different $\mathbf{V}_j$ are orthogonal by the definition of $\mathbf{V}$. Hence $\mathbf{\Psi}$ has power Pe$^2$ for any real vector $\mathbf{a}$. When 
$\|\mathbf{a}\| = O(\epsilon)$, the flow (\ref{Psi}) is a discrete version of the flow (\ref{psipert}). In this case we can write a Taylor expansion for Nu like (\ref{Nuexpn}) but in powers of $\mathbf{a}$:
\begin{align}
    \mbox{Nu}(\mathbf{a}) = \mbox{Nu}(\mathbf{0}) +  \mathbf{a}^T D\mbox{Nu}\bigg|_{\mathbf{a} = \mathbf{0}}+\frac{1}{2}\mathbf{a}^T D^2\mbox{Nu}\bigg|_{\mathbf{a} = \mathbf{0}} \mathbf{a} + \ldots. \label{Nua}
\end{align}
where $\mbox{Nu}(\mathbf{0})$ = Nu$_0$ is Nu for the steady flow and $D\mbox{Nu}$ and $D^2\mbox{Nu}$ are the gradient vector and Hessian matrix of first and second derivatives of Nu with respect to components of $\mathbf{a}$, respectively. We compute the gradient and Hessian as the limits of finite differences. The gradient entries are
\begin{align}
   D\mbox{Nu}_i \bigg|_{\mathbf{a} = \mathbf{0}} = \frac{\partial \mbox{Nu}}{\partial \mathrm{a}_i}\bigg|_{\mathbf{a} = \mathbf{0}} = \lim_{\epsilon \to 0} \frac{\mbox{Nu}(\epsilon\mathbf{e}_i)-\mbox{Nu}(\mathbf{0})}{\epsilon} = \mbox{Nu}_1(\epsilon\mathbf{e}_i)-\mbox{Nu}_1(\mathbf{0}) = 0. \label{DNu}
\end{align}
The limit of the finite difference quotient in (\ref{DNu}) is
computed by inserting the perturbation expansion of Nu (\ref{Nuexpn}) for the flows corresponding to $\mathbf{a} =\epsilon\mathbf{e}_i$ and 
$\mathbf{a} = \mathbf{0}$. As explained below (\ref{T1B}), Nu$_1$ = 0 for $\mathbf{a} =\epsilon\mathbf{e}_i$ for all $i$, since the first-order temperature field has zero time-average.
We also have Nu$_1$ = 0 for $\mathbf{a} =\mathbf{0}$ since 
Nu$(\mathbf{0}) \equiv$ Nu$_0$ so the remaining terms in (\ref{Nuexpn}) vanish. 
The entries of the Hessian matrix are also computed using the expansion (\ref{Nuexpn}):
\begin{align}
     D^2\mbox{Nu}_{ij} \bigg|_{\mathbf{a} = \mathbf{0}} &= \frac{\partial^2 \mbox{Nu}}{\partial \mathrm{a}_i\partial \mathrm{a}_j}\bigg|_{\mathbf{a} = \mathbf{0}} = \lim_{\epsilon \to 0} \frac{\mbox{Nu}(\epsilon\mathbf{e}_i\!+\!\epsilon\mathbf{e}_j)-\mbox{Nu}(\epsilon\mathbf{e}_i)-\mbox{Nu}(\epsilon\mathbf{e}_j)+\mbox{Nu}(\mathbf{0})}{\epsilon^2} \label{Hessian}\\
    &= \mbox{Nu}_2(\epsilon\mathbf{e}_i\!+\!\epsilon\mathbf{e}_j)-\mbox{Nu}_2(\epsilon\mathbf{e}_i)-\mbox{Nu}_2(\epsilon\mathbf{e}_j). \label{Nu23}
\end{align}
These entries are nonzero in general, so the quadratic term in (\ref{Nuexpn}) is the leading order change to Nu due to the unsteady flow perturbation. 
\textcolor{black}{Thus, for a given (small) magnitude of $\|\mathbf{a}\|_2$, $\eta$ say, we maximize the quadratic term,
\begin{align}
\max_{\|\mathbf{a}\|_2 = \eta}
\frac{1}{2}\mathbf{a}^T D^2\mbox{Nu}\bigg|_{\mathbf{a} = \mathbf{0}} \mathbf{a}.
\end{align}
The optimal flow perturbation $\mathbf{a}$ is the eigenvector of the Hessian with the largest eigenvalue, which is real since the Hessian is symmetric. Therefore, we compute the Hessian for a range of $\tau$ and Pe, compute the largest eigenvalues in each case, and examine the corresponding eigenvectors (i.e flows). We will examine not only the flow corresponding to the largest eigenvalue but also the flows corresponding to other positive eigenvalues, to consider a wider range of beneficial flows.
}

For flows of the form (\ref{Psi}), the Hessian is 2$N_m$-by-2$N_m$, with $N_m = (2M+1)(N-3)$. The number of entries in each Hessian ranges from $ \approx 6 \times 10^6$ for $n = 256$ to $\approx 10^8$ for $n = 1024$, reduced by a factor of 4 by symmetry. We compute the Hessian at a large number of $\tau$ and Pe values, so an efficient method is used to compute a single Hessian matrix, i.e. the large number of entries given by (\ref{Nu23}) for all $i$ and $j$. The computing time ranges from 1-2 hours for $n$ = 256 to about 8 days for $n$ = 1024 on a single processor.  

The sequence of computations for the Hessian is described in detail in \cite{alben2024enhancing}, so here we give a summary.
First, given $\psi_0$, we
solve the zeroth-order problem (\ref{T0}) for $T_0$. Next, we compute Nu$_2(\epsilon\mathbf{e}_i)$ for $i = 1,\ldots, N_m$ by setting the flow mode coefficient vector
$\mathbf{a} = \epsilon\mathbf{e}_i$ for $i = 1,\ldots, N_m$. Therefore in (\ref{psipert}) we 
set $f_A$ to column $i$ of $\mathbf{V}$, $f_B$ to $\mathbf{0}$, and $c_1$ to 1. Then we follow the remaining steps in that section to get Nu$_2$: we solve the system (\ref{T1A})--(\ref{T1B}) for ($T_{1A}$, $T_{1B}$) and then solve (\ref{T2s}) for $T_{2s}$ and compute Nu$_2$ from (\ref{Nu2}). To compute Nu$_2(\epsilon\mathbf{e}_i)$ for $i = N_m+1,\ldots, 2N_m$, the steps are the same but we now have a single sine term instead of a cosine term in (\ref{Psi}). So we set $f_A$ to $\mathbf{0}$, $f_B$ to column $i$ of $\mathbf{V}$, and $c_1$ to 1.
Having computed Nu$_2(\epsilon\mathbf{e}_i)$ in (\ref{Nu2}) for all $i$, the same set of values gives Nu$_2(\epsilon\mathbf{e}_j)$ for all $j$.

For Nu$_2(\epsilon\mathbf{e}_i\!+\!\epsilon\mathbf{e}_j)$, we have two unsteady terms in (\ref{Psi}) which may appear in either the cosine or sine summand. We put this flow in the form (\ref{psipert}) by setting $c_1 = 2$ and setting $f_A$ to the sum of the columns of $\mathbf{V}$ (if any) that appear in the cosine summand and $f_B$ to the sum of the columns of $\mathbf{V}$ (if any) that appear in the sine summand. 
($T_{1A}$, $T_{1B}$) is simply the sum of those for Nu$_2(\epsilon\mathbf{e}_i)$ and Nu$_2(\epsilon\mathbf{e}_j)$, since the right-hand side of (\ref{T1A})--(\ref{T1B}) is linear in $f_A$ and $f_B$. We then form the right-hand side of (\ref{T2s}), solve it for $T_{2s}$, and compute Nu$_2$ from (\ref{Nu2}).

Actually, we can obtain Nu$_2$ for each right-hand side of (\ref{T2s}) without having to solve the matrix equation. In discretized form, equations (\ref{T2s})--(\ref{Nu2}) are a linear mapping from the right-hand-side vector of (\ref{T2s}) to the scalar Nu$_2$, and this mapping is given by an inner product of a fixed vector with the right-hand side of (\ref{T2s}), as explained in \cite{alben2024enhancing}. So the dominant cost is simply forming the $\sim 10^7$--$10^8$ right-hand-side vectors, which may still take several days on a single CPU due to the large number of them.

We mentioned that only about 1/4 of the Hessian entries need to be computed by symmetry. We show in appendix A of \cite{alben2024enhancing} that the Hessian is block skew-symmetric:
\begin{align}
    D^2\mbox{Nu} =\left[\begin{array}{cc}
A &B \\ -B&A \\ 
 \end{array}\right] \label{BlockSkewSymm}
\end{align}
and since the Hessian is also symmetric, the blocks $A$ and $B$ are symmetric and skew-symmetric, respectively. Therefore to form the Hessian we only need to compute about half the entries of $A$ and $B$, or about one quarter of the entries of the Hessian.

We have described how the Hessian is computed for flows of the form (\ref{Psi}), with just a single period.
If the flow has unsteady perturbation modes with multiple periods, we show in appendix B of \cite{alben2024enhancing} that the Hessian is block diagonal, with the blocks consisting of the Hessians for each period alone.
Therefore, the problem is decoupled, with 
the eigenvalues and eigenvectors given by those for each period alone.  Next, we compute the single-period blocks of the Hessian for various values of the period, and the corresponding eigenvectors and eigenvalues. We focus on cases with positive eigenvalues, so the eigenvectors give flow perturbations that increase Nu.

\section{Eigenvalue distributions and eigenvectors (flow modes) \label{sec:eval}}

\begin{figure}
    \centering
    \includegraphics[width=1\textwidth]{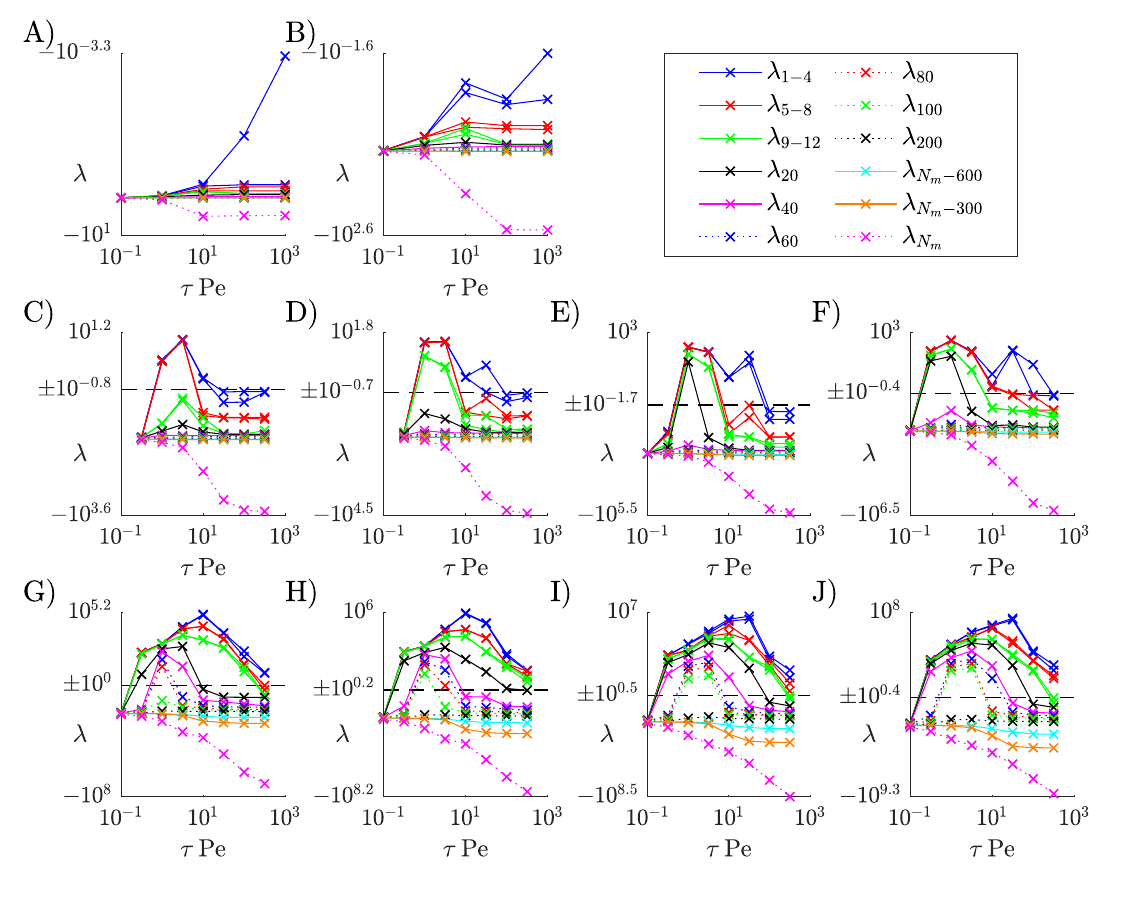}
    \caption{\footnotesize Distributions of Hessian eigenvalues at Pe = 10$^2$ (panel A), 10$^3$ (B), 10$^{3.5}$ (C),
    10$^4$ (D), 10$^{4.5}$ (E), 10$^5$ (F), 10$^{5.5}$ (G), 10$^6$ (H), 10$^{6.5}$ (I), and 10$^7$ (J).} 
    \label{fig:WallToWallEigenvalueDistributionsFig}
             \vspace{-.10in}
\end{figure}

In figure \ref{fig:WallToWallEigenvalueDistributionsFig} we plot the distributions of eigenvalues for a range of Pe, from 10$^2$ to 10$^7$ (one value per panel).
Within each panel, we plot the eigenvalue distributions across ranges of $\tau$Pe, from 10$^{-1}$ to 10$^{2.5}$, where the largest eigenvalues occur. We use $\tau$Pe instead of $\tau$ because we find that the peak eigenvalues occur at $\tau$Pe $\in [10^0, 10^{1.5}]$, so the optimal $\tau$ seems to scale approximately as Pe$^{-1}$, as was found for channel flows also \cite{alben2024enhancing}. We will discuss this scaling further shortly. In each panel, across $\tau$Pe, we plot selected eigenvalues ranging from the largest ($\lambda_1$) to the smallest ($\lambda_{N_m}$). The eigenvalues occur in identical pairs due to the block skew-symmetry of the Hessian (\ref{BlockSkewSymm}), as explained in \cite{alben2024enhancing}. Physically, the pairs correspond to a phase shift by 1/4 period in time, which takes the ($\cos(2\pi t/\tau)$, $\sin(2\pi t/\tau)$) components of a flow to ($\sin(2\pi t/\tau)$, -$\cos(2\pi t/\tau)$) components but results in the same Nu.

In some cases the pairs approximately coalesce into a group or four (or more) nearly identical values. In figure \ref{fig:WallToWallEigenvalueDistributionsFig} the leading three quartets of eigenvalues are shown with solid blue, red, and green lines respectively. In some cases (The right sides of panels A--F) two distinct blue lines can be seen, corresponding to two distinct pairs of values. In other cases, (left sides of panels A--F), the two blue lines overlap, sometimes together with the two red lines. These coalescences into groups of four or more correspond to symmetric groups of flows for the associated eigenvectors. In other words, the eigenvectors' flow fields map into each other under reflections about the horizontal and/or vertical midlines of the domain, corresponding to the symmetries of the steady flows in figure \ref{fig:StaticFlowsFig}A--H. We will give concrete examples of such flows after discussing this figure.

After the leading three quartets, lines are given in figure \ref{fig:WallToWallEigenvalueDistributionsFig} for every 20th eigenvalue up to $\lambda_{100}$, then $\lambda_{200}$, and the last three lines give a sense of the eigenvalue distribution near the bottom of the spectrum. These 15 lines (including 2 for each of the three leading quartets) give a sense of the overall distribution of the $N_m \approx$ 2400--10,000 eigenvalues. At Pe = 10$^2$ (panel A) and 10$^3$ (panel B), all eigenvalues are negative. At Pe = 10$^{3.5}$ (panel C), a band of positive eigenvalues appears. The dashed line in this and subsequent panels marks a range of values near zero which is omitted from the plot. This omission allows us to give one logarithmic scale for positive eigenvalues (above the dashed line) and another for negative eigenvalues (below the dashed line). 
The range of omitted values is chosen to be small enough to exclude all the values on the 15 lines. The peak eigenvalue in panel C occurs at $\tau$Pe = 10$^{0.5}$. A second peak occurs in panels D--F at $\tau$Pe = 10$^{1.5}$, followed by a single peak in panels G--J, at the Pe with branched steady flows. 
As Pe increases from panel C to J, both the number of positive eigenvalues at each $\tau$Pe and their maxima increase. These last two properties are also shared by the eigenvalues in the channel flow problem \cite{alben2024enhancing}. There the eigenvalue distributions are simpler and smoother, with a single peak that is more symmetrical about its maximum. This may be related to the simpler unidirectional form of the base steady channel flows.

Most of the eigenvalues lie between $\lambda_{200}$ and $\lambda_{N_m-600}$, the black dotted lines and cyan solid lines. These are relatively close together on the negative part of the vertical axis, so there are no more than 200 positive eigenvalues in all these cases, out of $N_m \approx 5000$ in C--F and 10,000 in G--J. At the bottom of the range, the eigenvalues decrease sharply from $\lambda_{N_m-600}$ to $\lambda_{N_m}$, at $\tau$Pe $\gtrapprox 10^0$. 

\begin{figure}
    \centering
\includegraphics[width=0.93\textwidth]{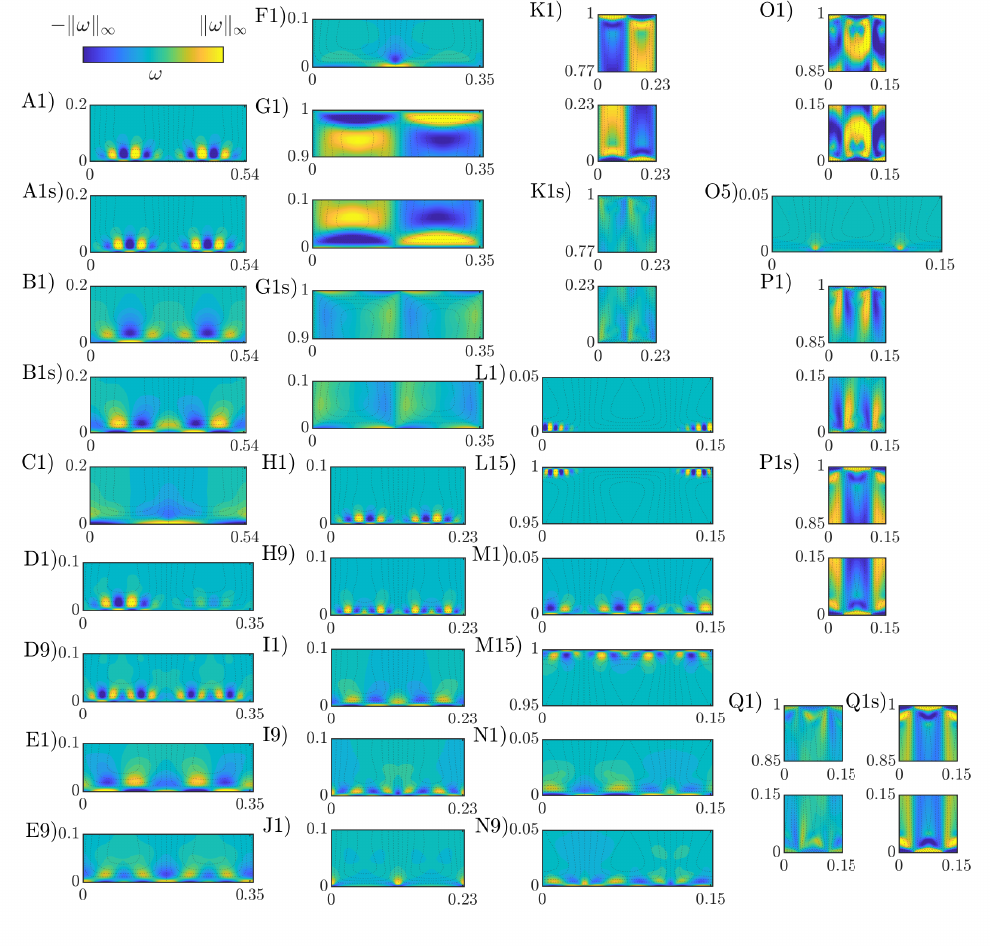}
    \caption{\footnotesize Vorticity fields for a selection of leading eigenmodes at Pe = $10^{3.5}$--$10^5$, when the steady base flows are unbranched. Each panel is labeled with a letter that corresponds to Pe and $\tau$Pe. Following the letter is the mode number (in order from largest to smallest eigenvalue). In a few cases the number is followed by ``s," denoting the $\sin(2\pi t/\tau)$ component. All other cases show the $\cos(2\pi t/\tau)$ component. Letters A--C show eigenmodes for Pe = 10$^{3.5}$ and $\tau$Pe = 10$^{0}$, 10$^{0.5}$, and 10$^{1}$, respectively. Letters D--G show eigenmodes for Pe = 10$^{4}$ and $\tau$Pe = 10$^{0}$, 10$^{0.5}$, 10$^{1}$, and 10$^{1.5}$, respectively. Letters H--K show eigenmodes for Pe = 10$^{4.5}$ and $\tau$Pe = 10$^{0}$, 10$^{0.5}$, 10$^{1}$, and 10$^{1.5}$, respectively. Letters L--Q show eigenmodes for Pe = 10$^{5}$ and $\tau$Pe = 10$^{-0.5}$, 10$^{0}$, 10$^{0.5}$, 10$^{1}$, 10$^{1.5}$, and 10$^{2}$, respectively.} 
\label{fig:EigVectsFigLowPe}
             \vspace{-.10in}
\end{figure}

We now show the corresponding eigenvectors, and only a small selection of those with positive eigenvalues. Even in this small set we find a range of patterns. Figure \ref{fig:EigVectsFigLowPe} shows the vorticity fields for eigenvectors with Pe = $10^{3.5}$--$10^5$, corresponding the unbranched steady flows (figure \ref{fig:StaticFlowsFig}D--G). Each vorticity field is labeled with a letter corresponding to the Pe and $\tau$Pe values, followed by the mode number (1 denoting the mode with largest eigenvalue), and then in a few cases ``s" for the $\sin(2\pi t/\tau)$ component. All cases without ``s" show the $\cos(2\pi t/\tau)$ component. Letters A--C show modes at Pe = $10^{3.5}$ with $\tau$Pe = 10$^{0}$, 10$^{0.5}$, and 10$^{1}$, respectively. 
Pe = $10^{3.5}$ is the lowest Pe in figure \ref{fig:WallToWallEigenvalueDistributionsFig} with positive eigenvalues. The flows consist of chains of vortices of alternating sign near the walls, very similar to those for the channel flows in \cite{alben2024enhancing} at small to moderate $\tau$Pe. In that work the top modes consisted of a single vortex chain extending along the channel wall. Panel A1 here shows the $\cos(2\pi t/\tau)$ component of the top mode, which has two vortex chains, corresponding to the two convection rolls of the steady flow with streamlines shown by black dotted lines. The steady flow is rightward and leftward along the left and right sides of the bottom wall, respectively. Panel A1s shows the $\sin(2\pi t/\tau)$ component of the top mode, which also has two vortex chains, but shifted spatially in phase from those of the cosine mode by 1/4 of the spatial period of the vortex chain. When the cosine and sine components are superposed, the spatial and temporal shifts result in two counterpropagating traveling waves of vorticity moving at the speed of the steady flow near the vortices, as explained in \cite{alben2024enhancing} for the case of a single traveling wave. Movie 1 in the supplementary material shows the same phenomenon at a higher Pe, 10$^{4.5}$. Panels B1 and B1s show the corresponding modes with $\tau$Pe increased from 10$^{0}$ to 10$^{0.5}$. The vorticity field is similar to A1-A1s but the vortices are fewer and larger. The enlargement of vortices with increasing $\tau$Pe was also seen in \cite{alben2024enhancing}. Panel C1, with $\tau$Pe increased to 10$^1$, shows vorticity regions enlarged to half the horizontal domain size. This could be seen both as a continuation of the trend of larger discrete vortices and as part of a common transition to different vorticity patterns (usually more global) at 
$\tau$Pe = 10$^1$ and larger. This transition will be seen in the next groups of flows at higher Pe. 

The vortex chains are at the bottom walls in panels A--C. These eigenvectors actually occur in pairs (not shown), with the same vortex chains at either the top or bottom walls, and the same eigenvalues. Considering also the symmetry of the cosine and sine components of the eigenvectors mentioned earlier, we actually have quartets of identical eigenvalues. 

Moving on to the next larger Pe, 10$^4$, we now have positive eigenvalues at four $\tau$Pe, with modes shown in panels D--G. In D and E we show both the first and ninth modes, the latter representing the third quartet of modes, those with the smallest positive eigenvalues here. In D9 the vortex chains are broken into two clusters each on the left and right halves, similarly to the eigenmodes below the top mode for the channel flows of \cite{alben2024enhancing}. In panel F1, at $\tau$Pe = 10$^1$, we have the aforementioned transition to different vorticity patterns---here a pointlike vortex at the middle of the domain. At $\tau$Pe = 10$^{1.5}$, vortices fill the domain both horizontally and vertically in the cosine component (G1), with a rather different and weaker global vorticity pattern in the sine component (G1s). The larger vortices of panels G1--G1s mix the boundary layer fluid more strongly with fluid in the bulk.

The sequence is generally similar at Pe = 10$^{4.5}$, shown in panels H--K. Vortex chains occur in H--I, at smaller $\tau$Pe, and then a pointlike vortex in J1 is followed by domain-filling vortices in K1--K1s. This pattern is seen again at Pe = 10$^{5}$, panels L--Q. Vortex chains are seen in panels L--N, with $\tau$Pe $\leq$ 10$^{0.5}$. In panels O--Q, with $\tau$Pe $\geq$ 10$^{1}$, we have vertically elongated vorticity patterns together with pointlike vortices in panel O5. In O--Q the vertically elongated vorticity is shifted from the steady convection rolls by 1/4 of the horizontal period. These unsteady vortices thus transport fluid from one steady convection roll to the other.
We note that
the change of eigenmodes from vortex chains to other vorticity distributions at $\tau$Pe = 10$^{1}$ for all Pe in figure \ref{fig:EigVectsFigLowPe} coincides with the 
local minimum of the top two quartets of eigenvalues in figure \ref{fig:WallToWallEigenvalueDistributionsFig}D--F.

\begin{figure}
    \centering
\includegraphics[width=0.95\textwidth]{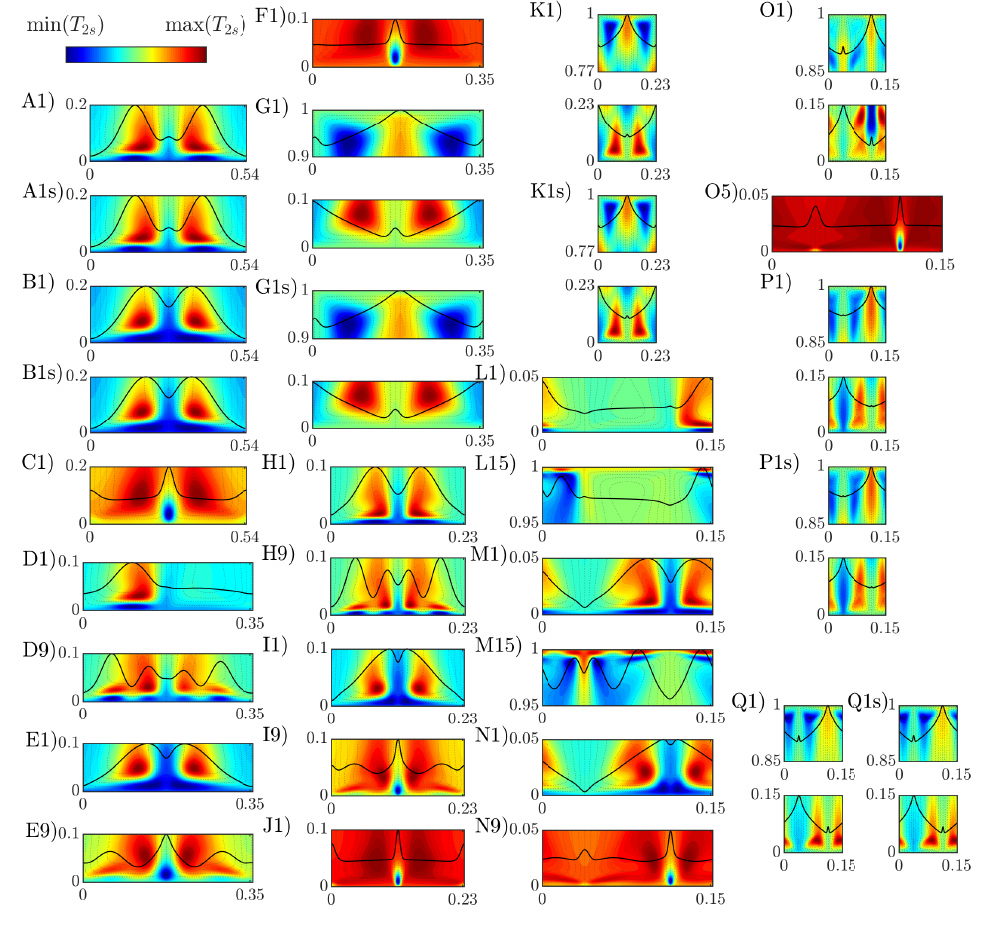}
    \caption{\footnotesize Perturbations in temperature fields ($T_{2s}$) and heat flux distributions ($-\partial_yT_{2s}$) at leading (quadratic) order for the flow modes in figure \ref{fig:EigVectsFigLowPe}.
    The black lines in each panel are graphs of $-\partial_yT_{2s}$ versus $x$ for the wall shown (top or bottom). These graphs are scaled vertically to fit within the panels, with the value 0 at the vertical midpoint of each panel.} 
\label{fig:T2sFigLowPe}
             \vspace{-.10in}
\end{figure}

In the next section, we will consider the effect of the unsteady flow perturbations on heat transfer, for perturbation amplitudes ranging from small to large. First, we examine the leading-order (quadratic) effect on the time-averaged temperature and on Nu, in the limit of small perturbations. In figure \ref{fig:T2sFigLowPe}, the color fields show $T_{2s}$ for the flow perturbations in figure \ref{fig:EigVectsFigLowPe}, in the same regions of space. In each panel, the black line is a graph of the corresponding heat flux out of the bottom wall or into the top wall, with the value 0 at the vertical midpoint of each panel. For both top and bottom walls, enhanced heat transfer occurs where the black line lies above the midpoint, i.e. $-\partial_yT_{2s} > 0$. 

In panels A1--B1s, the black graphs show two main peaks in the perturbation's average heat flux from the bottom wall $-\partial_yT_{2s}$. The peaks lie above dark blue regions in $T_{2s}$ near the lower (hot) wall, showing increased cooling there. These are roughly where the vortex chains lie in figure \ref{fig:EigVectsFigLowPe}A1--B1s. The enhancement is partly canceled by the black curves' troughs near the periodic boundary. The black curves have similar peaks near the vortex chains in panels D1, E1, H1, I1, L1, M1, and N1, though the curves differ in some respects, e.g. the troughs' shapes. In general the troughs correspond to regions that the vortex chains move away from, as seen in movies 1--4. The heat flux perturbations for pointlike vortices (F1, J1, and O5, all with $\tau$Pe = 10$^1$) have a common form, different from those of the vortex chains. The heat flux peaks near the pointlike vortices and is nearly zero elsewhere. C1 also fits this pattern though the vortices are less pointlike. For the global vorticity distributions that occur at $\tau$Pe $>$ 10$^1$ (K1, P1, and Q1), the heat flux distributions typically have cusp-like peaks that lie between convection rolls of the perturbation flows.

\begin{figure}
    \centering
\includegraphics[width=0.95\textwidth]{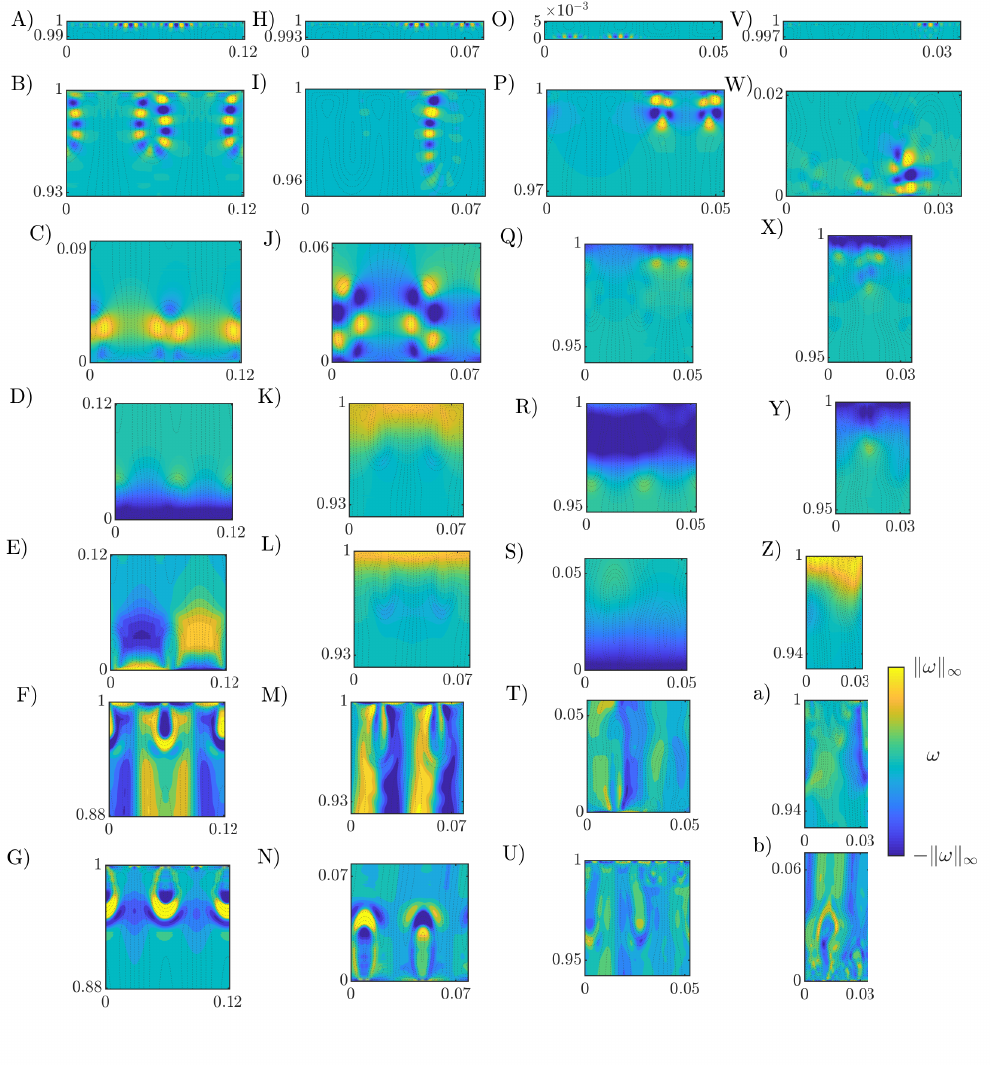}
    \caption{\footnotesize Vorticity fields for the cosine components of the top eigenmodes at Pe = $10^{5.5}$--$10^7$, when the steady base flows are branched. From left to right, the four columns correspond to Pe = 10$^{5.5}$ (A--G), 10$^{6}$ (H--N), 10$^{6.5}$ (O--U), and 10$^{7}$ (V--Z, a, b). The seven panels in each column correspond to $\tau$Pe = 10$^{-0.5}$, 10$^{0}$, \ldots, 10$^{2.5}$, from top to bottom.} 
\label{fig:EigVectsFigHighPe}
             \vspace{-.10in}
\end{figure}

We now consider the top eigenmodes at Pe $\geq 10^{5.5}$, beyond the branching transition of the steady flows. The eigenvalue distributions are shown in the last four panels of figure \ref{fig:WallToWallEigenvalueDistributionsFig}, G--J. Now there are many positive eigenvalues at the seven $\tau$Pe from 10$^{-0.5}$ to 10$^{2.5}$. In figure \ref{fig:EigVectsFigHighPe} we present just the eigenmode of the largest eigenvalue, and just its cosine component, which is often similar to the sine component. The four columns from left to right show the top eigenmodes for Pe = 10$^{5.5}$, 10$^{6}$, 10$^{6.5}$, and 10$^{7}$. The seven rows from top to bottom show the top eigenmodes for $\tau$Pe = 10$^{-0.5}$, 10$^{0}$, \ldots, 10$^{2.5}$. At $\tau$Pe = 10$^{-0.5}$ (top row), the modes consist of vortex chains along the top or bottom wall. The vortices become ever smaller as Pe increases, but the main difference with the previous lower-Pe figure is that the vortex chains are shorter horizontally. In particular, the vortex chains avoid the regions on the wall adjacent to the enclosed eddies between the branches. In panel A these regions occur at the center and the two sides of the panel (seen more clearly by the black dotted streamlines in panel B, which are the same in panel A). In panel H the vortices are clustered on either side of the enclosed eddy on the right side (seen more clearly in panel I), and in panels O and V the eddies are concentrated on single flow branches next to enclosed eddies.  

The second row, with $\tau$Pe = 10$^{0}$, shows a feature unique to the branched flows---the vortex chains follow the streamlines off of the wall, along enclosed eddies near the wall. The sine components again have vortex chains that are shifted by 1/4 spatial period, so the vortex chains move as traveling waves around the eddies (see movie 5). The vortices in the vortex chains are larger at $\tau$Pe = 10$^{0.5}$ (third row), and then in the fourth row the individual vortices lose their circular shape and become spread out along the eddy, showing again that there is a transition at $\tau$Pe = 10$^{1}$. The fifth row is similar to the fourth. In the sixth and seventh rows, the vortices are more elongated and curve around the eddies, sometimes appearing in elongated pairs of opposite-signed vorticity. These features persist in the seventh row, but strong vorticity is also concentrated in pointlike regions near the wall that are difficult to see clearly. We have shown only the top eigenmodes but briefly comment on the qualitative features of the second through 16th eigenmodes. At smaller $\tau$Pe (the first four rows), the subsequent modes have vortex chains like the top eigenmodes, but they are placed at different eddies or branches of the steady flows. At larger $\tau$Pe, (the last three rows), the subsequent modes have similar elongated vortices superposed on large regions of nonzero vorticity that extend far from the walls.

\begin{figure}
    \centering
\includegraphics[width=0.95\textwidth]{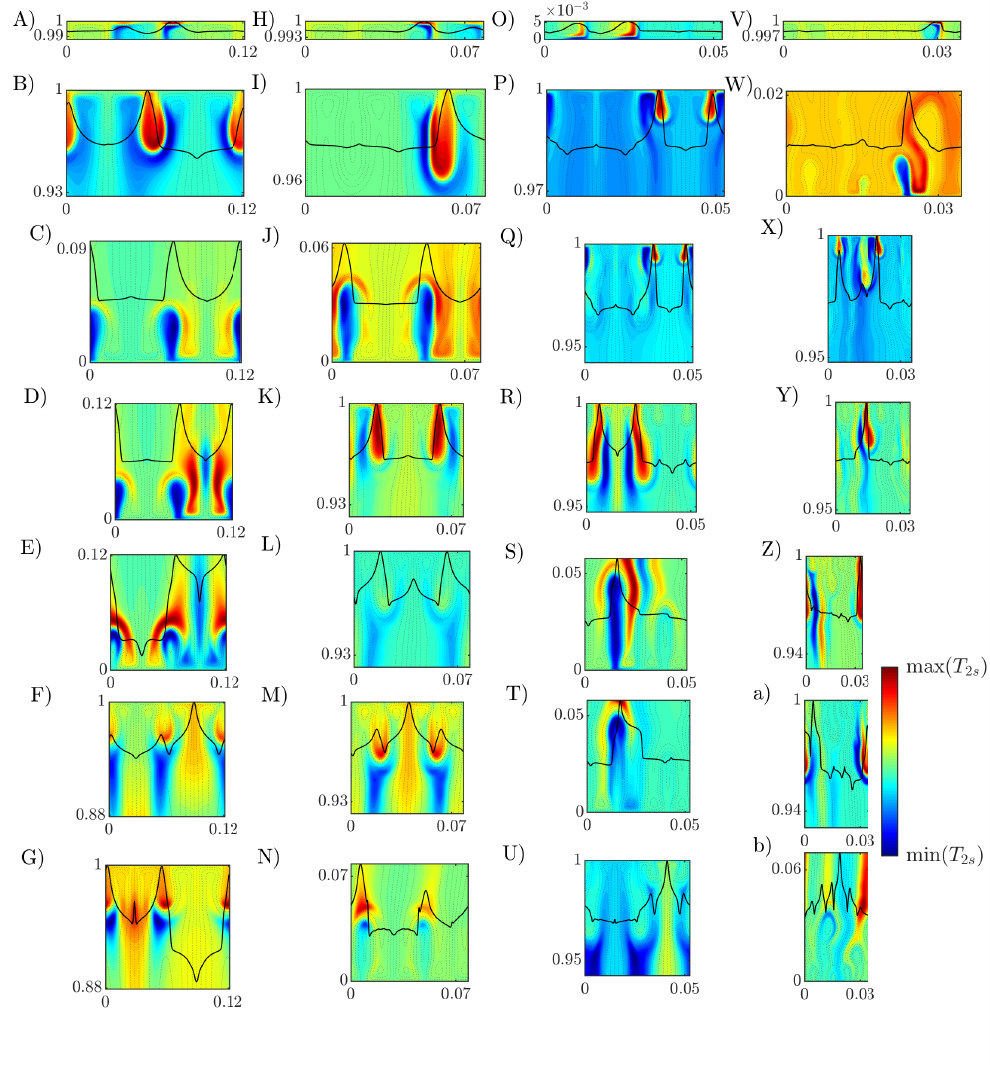}
    \caption{\footnotesize Perturbations in temperature fields ($T_{2s}$) and heat flux distributions ($-\partial_yT_{2s}$) at leading (quadratic) order for the flow modes in figure \ref{fig:EigVectsFigHighPe}.
    The black lines in each panel are graphs of $-\partial_yT_{2s}$ versus $x$ for the wall shown (top or bottom). These graphs are scaled vertically to fit within the panels, with the value 0 at the vertical midpoint of each panel.} 
\label{fig:T2sFigHighPe}
             \vspace{-.10in}
\end{figure}

Figure \ref{fig:T2sFigHighPe} shows the leading perturbations in the temperature fields ($T_{2s}$) and heat flux ($-\partial_yT_{2s}$, black curves)
corresponding to the flows in figure \ref{fig:EigVectsFigHighPe}. The top row shows peaks in the heat flux perturbation near the vortex chains along the wall, as before. The second and third rows also have peaks in heat flux perturbations near the vortex chains, even though the vortex chains now extend away from the walls while the heat flux occurs at the wall. The fourth and fifth rows show sharp heat flux peaks even though the corresponding vorticity in figure \ref{fig:EigVectsFigHighPe} is more spread out across the wall. However, there are faint opposite-signed vortices visible further from the wall--slightly yellow regions in panels D, R, S, and Y and slightly blue regions in panels K, L, and Z, and the heat flux peaks seem to approximately correspond to these regions both in number and horizontal location.
In the sixth and seventh rows the heat flux peaks generally correspond to regions of strong vorticity or interfaces between opposite signed vorticity, though the relationship is less clear, particularly in panels U and b.

Now that we have described the eigenvalue distributions and some of the top eigenmodes and the resulting heat flux patterns, we have some understanding of the optimal flow perturbations and how they increase heat flux. The flows share some features of the optimal perturbations of channel flows, such as vortex chains, but with additional complexity due to the structure of the steady optimal flows here, thin rectangular convection rolls with branching. The resulting heat flux distributions have peaks and troughs that generally correspond to features of the vorticity patterns.  

Next, we show the effect of the unsteady flow perturbations on the Nusselt number as we vary the perturbation magnitude $\epsilon$ from small to large. This requires a more capable unsteady advection-diffusion solver than the one used for channel flows in \cite{alben2024enhancing}, as we explain next.


\section{Unsteady simulations from small to large amplitude \label{sec:UnsteadySims}}

We now examine the performance of the optimal eigenmodes in the small-to-large amplitude regime. We solve the unsteady advection-diffusion equation (\ref{AdvDiff}) for flows of the form (\ref{psipert}), with $\psi_0$ set to one of the steady optima in figure \ref{fig:StaticFlowsFig}, $f_A$ and $f_B$ corresponding to one of the top eigenmodes, $c_1 = 1$, and the amplitude parameter $\epsilon$ ranging from $10^{-3}$ to $10^{-0.25}$. At the upper end of this range the unsteady perturbation is large, with energy close to that of the steady flow. We have cross-checked the simulations and eigenvalue computations, verifying that for small $\epsilon$, Nu $\approx$ Nu$_0$ + $\lambda_i \epsilon^2$/2 in agreement with the expansion (\ref{Nua}), when $\mathbf{a}$ is the Hessian eigenvector with norm $\epsilon$ corresponding to eigenvalue $\lambda_i$.
We primarily use the simulations to determine the range of $\epsilon$ for which the perturbations are effective, and characterize the resulting temperature fields.


On the finest grid (256-by-1024), the steady version of the discretized advection-diffusion equation can be solved in a few seconds using a direct solver (i.e. ``backslash" or mldivide in Matlab). In \cite{alben2024enhancing} we solved the unsteady advection-diffusion equation with Crank-Nicolson time-stepping for 20 periods of the unsteady flow, by which time the unsteady flows had usually reached a periodic steady state. With 200 time steps per period, 4000 time steps were required, with a cost equivalent to about 4000 steady solves. Unlike the channel flows, the wall-to-wall flows are still quite far from the periodic state after 20 periods unfortunately, particularly at Pe = 10$^6$ and 10$^7$ but also at smaller Pe such as 10$^4$. The difference may be due to underlying steady flows.
In \cite{alben2024enhancing} the steady flows are unidirectional flows through a channel, unlike the thin convection rolls of the wall-to-wall problem that have fine eddies and branched structures. For the channel flows, fluid parcels with initial deviations from the periodic steady-state temperature solution are rapidly advected out of the channel, while in the wall-to-wall flows, they circulate within the flow domain and may cause more extensive and long-lasting temperature deviations throughout the domain. 

Because of the very slow convergence of the time-stepping approach, we instead solve the larger linear system for the temperature field that is fully coupled in time over the flow period. A space-time multigrid method has been developed for the time-periodic diffusion equation \cite{hackbusch1981fast} and used to solve an advection-diffusion problem for channel flow at Pe = O(10$^2$--10$^4$) \cite{xiaolinwang2014thesis}. The latter problem had a smoother flow field  which allowed for a uniform spatial grid. Such methods can display poor convergence for strongly advection-dominated problems \cite{de2025multigrid}, and are complicated to implement with nonuniform grids, so we adopt a simpler approach here.

\begin{figure}
    \centering
\includegraphics[width=0.8\textwidth]{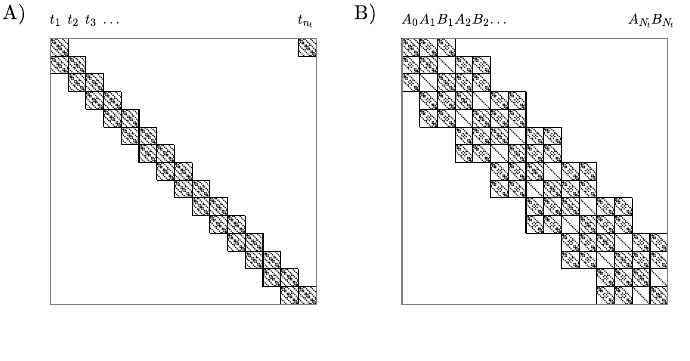}
    \caption{\footnotesize Matrix sparsity patterns for Crank-Nicolson discretization (A) and time-spectral method (B).} 
\label{fig:BlockMatricesFigure}
             \vspace{-.10in}
\end{figure}

First we consider the Crank-Nicolson second-order finite-difference time discretization of the advection-diffusion equation (\ref{AdvDiff}):
\begin{align}
    \left(1 + \frac{\Delta t}{2}\left(u\partial_x+v\partial_y-\nabla^2\right)\right)T^{j+1} = \left(1 - \frac{\Delta t}{2}\left(u\partial_x+v\partial_y-\nabla^2\right)\right)T^{j}, \quad j = 1, \ldots, n_t
\end{align}
where $T^j(x,y)$ is the temperature field at time step $j$, which ranges over the $n_t$ time steps of a period. Periodicity implies $T^{n_t+1}\equiv T^{1}$, resulting in a closed system of equations for $\{T^{1},\ldots, T^{n_t}\}$. With the unknowns ordered first with increasing $x$ at a given $y$ and $t$, then by increasing $y$ at a given $t$, then by $t$, the sparsity pattern of the linear system is shown in figure \ref{fig:BlockMatricesFigure}A. Each block row and block column correspond to the equations and unknowns at certain time steps, respectively.

The second approach we consider is the time-spectral method. We expand the temperature field as
\begin{align}
    T(x,y,t) = 1-y+\sum_{k = 0}^{N_t} A_k(x,y)\cos\left(\frac{2\pi k t}{\tau}\right) + \sum_{k = 1}^{N_t} B_k(x,y)\sin\left(\frac{2\pi k t}{\tau}\right). \label{TexpnNt}
\end{align}
The velocity field corresponding to (\ref{psipert}) is of the form
\begin{align}
u(x,y,t) &= u_{St}(x,y) + u_A(x,y) \cos\left(\frac{2\pi t}{\tau}\right) + u_B(x,y) \sin\left(\frac{2\pi t}{\tau}\right) \nonumber\\ 
v(x,y,t) &= v_{St}(x,y) + v_A(x,y) \cos\left(\frac{2\pi t}{\tau}\right) + v_B(x,y) \sin\left(\frac{2\pi t}{\tau}\right). \label{uv}
\end{align}
We insert (\ref{TexpnNt}) and (\ref{uv}) into (\ref{AdvDiff}) and set to zero the inner products of the residual with cosines and sines with frequencies up to $2\pi N_t/\tau$ (i.e., the Galerkin method). When the unknowns are ordered $\{A_0, A_1, B_1, \ldots, A_{N_t}, B_{N_t}\}$, with the corresponding ordering of the equations for the modes, the linear system is banded, with the sparsity pattern shown in figure \ref{fig:BlockMatricesFigure}B. The products of sines and cosines in $u\partial_xT$ and $v\partial_yT$ introduce a coupling of $\{A_k,B_k\}$ with 
$\{A_{k-1},B_{k-1}\}$ and 
$\{A_{k+1},B_{k+1}\}$. The couplings between $A_k$ and $B_{k+1}$ and $B_k$ and $A_{k-1}$ result in the blocks farthest from the main diagonal, on the third block diagonals above and below it respectively, giving 7 diagonals of blocks.

\begin{table}
\centering
\begin{subfloat}[Crank-Nicolson run time data.]{
\renewcommand{\arraystretch}{1} 
\setlength{\tabcolsep}{0.2cm} 
\begin{tabular}{c|c|c}
$n_t$ &  peak RAM (Gb) & run time (sec.) \\ \hline
& &\\[-.3cm]
4 & 4.6  & 12.7\\[.05cm]
5 & 7.1 & 21.2\\[.05cm]
6 & 12.2 & 33.8\\
7 & 14.8 & 45.3\\[.05cm]
8 & 20.8 & 77\\[.05cm]
9 & 22.2$^*$ & 185\\
\end{tabular}}
\end{subfloat}
\hspace{1.7cm}
\begin{subfloat}[Time spectral run time data.]{
\renewcommand{\arraystretch}{1} 
\setlength{\tabcolsep}{0.2cm} 
\begin{tabular}{c|c|c}
$N_t$ &  peak RAM (Gb) & run time (sec.) \\ \hline
& &\\[-.3cm]
3 & 4.9  & 4.8\\[.05cm]
5 & 8.7 & 16.8\\[.05cm]
7 & 15.8 & 32.9\\
9 & 21.9$^*$ & 153\\[.05cm]
\end{tabular}}
\end{subfloat}\\
\caption{Peak RAM usage and total run time to solve a) the Crank-Nicolson linear system (figure \ref{fig:BlockMatricesFigure}A) for various $n_t$ and b) the time spectral linear system (figure \ref{fig:BlockMatricesFigure}B) for various $N_t$. The spatial grid parameters are  $m = n = 256$.} \label{tab:RAMTime}
\end{table}

In table \ref{tab:RAMTime}(a) and (b) we show the peak memory (RAM) used and the total run time to solve the linear systems in figure \ref{fig:BlockMatricesFigure}A and B, respectively, using Matlab's backslash. We use $m = n = 256$, the coarsest spatial mesh, which is used at the smallest Pe, $\leq 10^4$. The cost is much higher when $n$ is increased to 512 and 1024. In the last line of each table, the peak RAM is starred because the solver was close to the maximum RAM available and therefore virtual memory, a much slower form of memory, was used. From this point the run times grow much more rapidly. 

Table \ref{tab:RAMTime} shows that both methods for the direct solution of the unsteady system require large memory, even for a small number of time steps per period $n_t$ (for Crank-Nicolson) and a small number of temporal frequencies $N_t$ (for the time-spectral method). This problem is well suited to the time-spectral method because the advection-diffusion equation (\ref{AdvDiff}) has a very smooth time dependence, through the flow velocities only, which have just the single frequency $2\pi/\tau$. In such cases exponential decay of $A_k$ and $B_k$ with $k$ in (\ref{TexpnNt}) is typical \cite{boyd2001chebyshev}. Furthermore, in the small-$\epsilon$ expansion for $T$ (\ref{TexpnEps}), the frequency $4\pi/\tau$ does not appear until the $\epsilon^2T_{2u}$ term, and higher frequencies appear at still higher powers of $\epsilon$.  Hence one may hope for good accuracy with a small value of $N_t$. 

We solve the time-spectral system with $N_t = 10$, i.e. 21 temporal modes,  which is still too large for a direct solution, considering that we take $n = 1024$ for Pe $\geq 10^{5.5}$ and we solve at $O(10^3)$ parameter combinations, varying Pe, $\tau$Pe, the eigenmode number, and $\epsilon$ simultaneously. Therefore, we use the GMRES iterative method, which requires a good preconditioner since the system is ill-conditioned. 
For $m = 256$, $n = 256$--1024, and various Pe and $\tau$Pe used here, the condition numbers of just the upper left block (which is the steady advection-diffusion matrix) and the upper left 3-by-3 array of blocks are $10^8$--$10^{10}$. 

We use a block-Gauss-Seidel preconditioner that approximates the matrix by its lower triangular and diagonal blocks, and then solves for the modes at each frequency sequentially from low to high. I.e., we solve the first block row of equations for $A_0$ assuming $A_1 = B_1 = 0$, then solve the next two block rows of equations simultaneously for $A_1$ and $B_1$  using the just-computed $A_0$, assuming $A_2 = B_2 = 0$, and so on. This approximation is particularly good at small $\epsilon$, where continuing the asymptotic calculation in section \ref{sec:pert} shows that the frequency-$j$ terms are $O(\epsilon^j)$. 

We run the iterative solver for $T$ at 8 values of Pe (10$^{3.5}$, 10$^{4}$, \ldots, 10$^{7}$), 8 values of $\tau$Pe (10$^{-1}$, 10$^{-0.5}$, \ldots, 10$^{2.5}$), 4 mode numbers (1, 5, 9, and 13), and 6 values of $\epsilon$ (10$^{-3}$, 10$^{-2}$, 10$^{-1}$, 10$^{-0.75}$, 10$^{-0.5}$, and 10$^{-0.25}$). The mode numbers are spaced apart by four to avoid repetition of flows, as the top eigenvalues and eigenmodes often occur in nearly identical quartets.

\begin{figure}
    \centering
\includegraphics[width=1\textwidth]{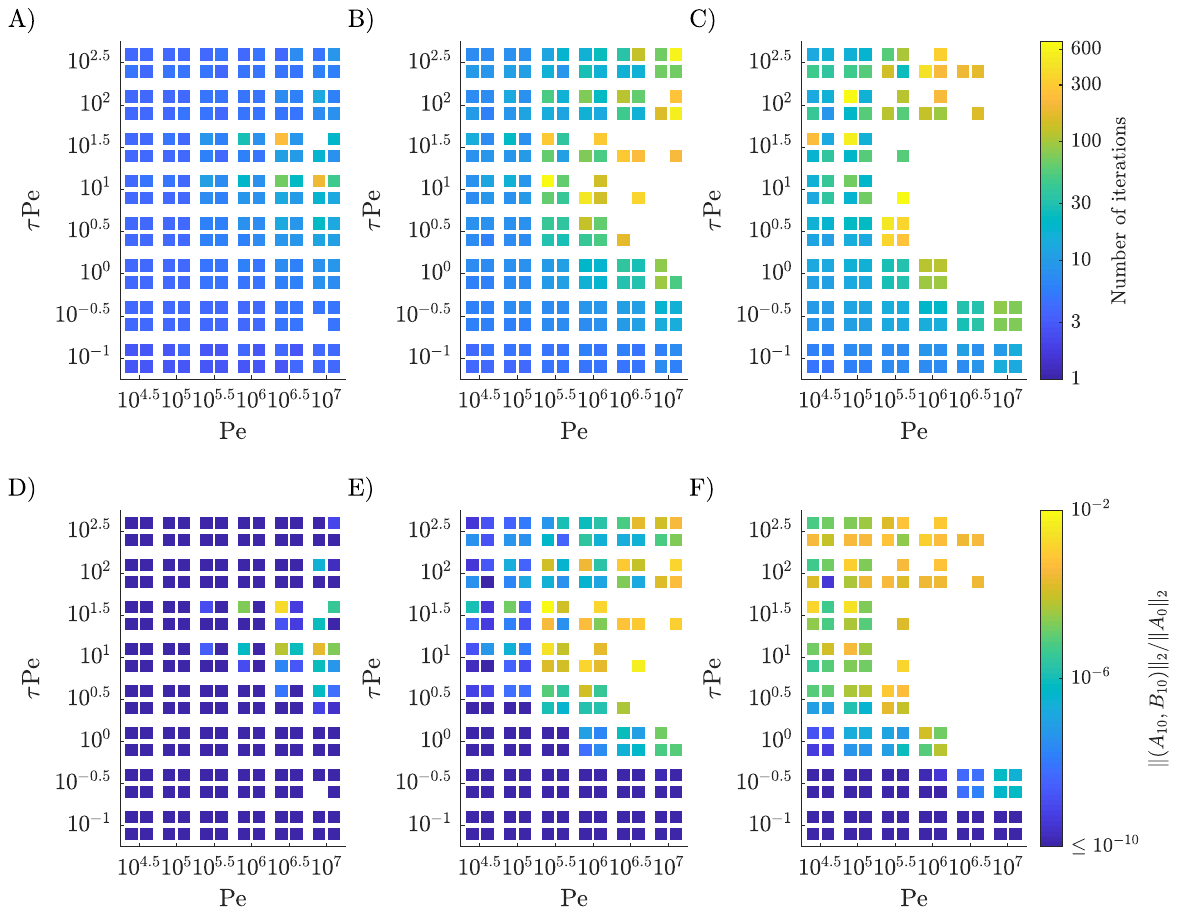}
    \caption{\footnotesize
    Number of iterations needed for the unsteady solver to converge to a relative residual of 10$^{-10}$ (A--C) and norm of highest wavenumber modes $\|(A_{10},B_{10})\|_2$ relative to the norm of the time-constant mode $\|A_0\|_2$ (D--F) at three $\epsilon$ values: $10^{-2}$ (A and D), $10^{-1}$ (B and E), and 10$^{-0.5}$ (C and F). The four squares clustered at each (Pe, $\tau$Pe) pair give the values for the four modes 1, 5, 9, and 13, at the upper left, upper right, lower left, and lower right squares in each cluster respectively.} 
\label{fig:NumItsRelModesFig}
             \vspace{-.10in}
\end{figure}

In some cases, particularly at higher Pe and higher $\epsilon$, the solver fails to converge within a few days of runtime, corresponding to about 700 iterations of preconditioned GMRES. The criterion for convergence is that the norm of the residual vector relative to the norm of the right-hand-side vector for the preconditioned system reaches 10$^{-10}$. The effect of parameters on convergence is shown in the top row of figure \ref{fig:NumItsRelModesFig}, which shows the number of iterations needed to converge. The perturbation amplitude $\epsilon = 10^{-2}$ (A), 10$^{-1}$ (B), and 10$^{-0.5}$ (C). The colored squares are missing where the iteration did not converge within a few days, most often at large Pe, where the condition number is larger, and at large $\epsilon$, where the unsteady components of the solution are more significant and the Fourier components $A_k$ and $B_k$ decay more slowly with $k$. This is shown in the second row, which plots the norm of the highest-frequency components of the temperature, $(A_{10}, B_{10})$ arranged as a vector, relative to the norm of the steady component $A_0$. Throughout panels D--F this relative norm is generally small, $\leq 10^{-2}$ but tends to be larger near the parameters where the iterations fail to converge. There is also a noticeable effect of $\tau$Pe on convergence. The largest number of failures occur at $\tau$Pe = $10^1$ and adjacent values, which happens to be where the flows transition from vortex chains to other vorticity patterns. We omit Pe = 10$^{3.5}$ and 10$^{4}$ on the horizontal axis because the pattern was similar to Pe = 10$^{4.5}$, i.e. relatively rapid convergence in most cases.   

Next, we discuss the Nu values for the cases that converged. For modes with positive eigenvalues, Nu increases above the steady value for some sequence of $\epsilon$ starting from 10$^{-3}$, then reaches a maximum, then decreases. In some cases the computations fail to converge before a decrease is seen.
For modes with negative eigenvalues, we find Nu decreases below the steady value as expected.

\begin{figure}
    \centering
\includegraphics[width=1\textwidth]{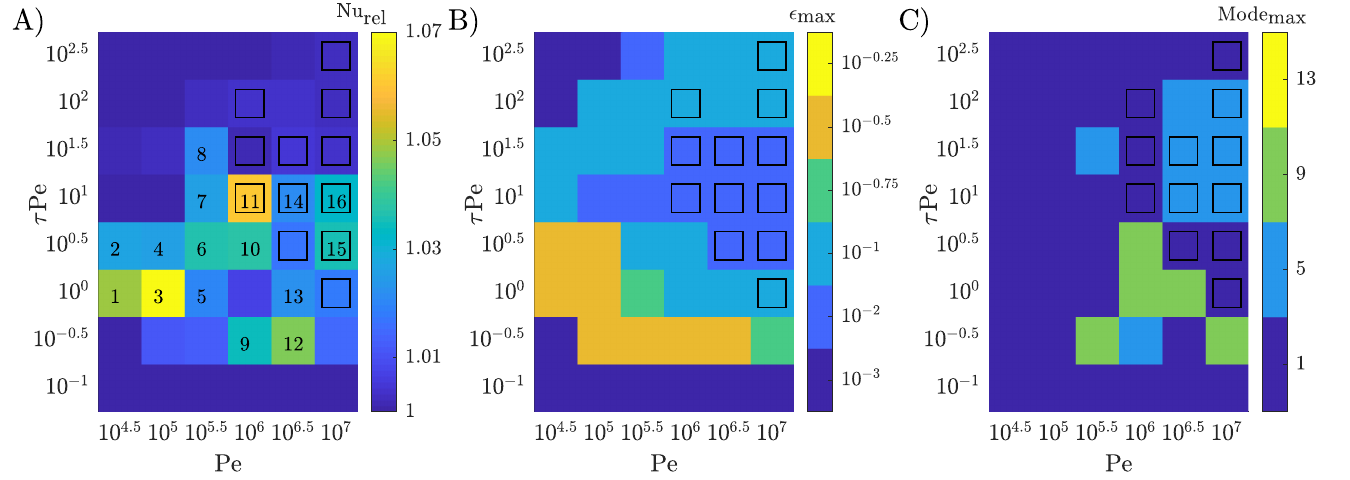}
    \caption{\footnotesize Properties of the unsteady perturbations that achieve the largest improvement in Nu over the steady optima, across ranges of Pe and $\tau$Pe. Panel A shows the factor of increase in Nu relative to the steady optimum at the same Pe. Panel B shows the value of $\epsilon$ at which the maximum increase in Nu occurs. Panel C shows which of the four tested modes achieves the maximum increase in Nu.} 
\label{fig:BestNuEpsModeFigure}
             \vspace{-.10in}
\end{figure}

Figure \ref{fig:BestNuEpsModeFigure} shows
the maximum Nu achieved by the unsteady flows, Nu$_\text{rel}$, which is Nu relative to that of the corresponding steady optimum.
Panel A shows the maximum value of Nu$_\text{rel}$ found at each Pe and $\tau$Pe. The increases in Nu$_\text{rel}$ above 1 (the steady value) were very small at Pe = 10$^{3.5}$, 10$^{4}$ so we restrict to larger Pe on the horizontal axis. The color of each box gives the maximum Nu$_\text{rel}$ over the cases that computed successfully out of the four mode numbers and six $\epsilon$ values at that
(Pe, $\tau$Pe) pair.
The maximum Nu$_\text{rel}$ of 1.07 occurs at Pe = 10$^5$ and $\tau$Pe = 10$^0$, and is significantly smaller than the maximum value of 1.56 seen for the channel flows \cite{alben2024enhancing}. Since the steady channel flows were constrained to be unidirectional, this difference would be reduced by allowing for fully 2D steady channel flows. Since the geometries and boundary conditions are different, some difference is expected in any case. Nu$_\text{rel}$ of 1.03 or higher mainly occur in the band of small-to-moderate $\tau$Pe = 10$^{-0.5}$--10$^{1}$, where the flows are a mixture of vortex chains and global vorticity patterns. A selection of cases with the largest Nu$_\text{rel}$ are numbered 1--16, and the vorticity and temperature fields for these cases are shown in the supplementary movies with the same number. In some colored boxes, mainly at larger Pe, a black square outline is shown, indicating that Nu$_\text{rel}$ occurred at the largest $\epsilon$ that computed successfully, so somewhat larger Nu$_\text{rel}$ might occur at larger $\epsilon$. For channel flows, by contrast, the largest Nu$_\text{rel}$ increased more strongly with Pe, and lack of convergence at large Pe was less significant. 

Panel B shows $\epsilon_\text{max}$, the $\epsilon$ at which the maximum Nu$_\text{rel}$ occurred. At $\tau$Pe = 10$^{-1}$, all eigenvalues are negative so the smallest $\epsilon$ is best. Just above this $\tau$Pe, a fairly large $\epsilon$ of 10$^{-0.5}$ is often best, though Nu$_\text{rel}$ is not much above 1. At $\tau$Pe $\geq 10^1$, smaller $\epsilon$ values, 10$^{-2}$ and 10$^{-1}$, are often best, though many cases at larger $\epsilon$ did not converge. 
Panel C shows the mode numbers that achieved the largest Nu$_\text{rel}$. At smaller Pe, the first mode is best. At larger Pe, the fifth mode is sometimes best at larger $\tau$Pe, while the ninth mode sometimes is best at smaller $\tau$Pe. As occurred for the channel flows, the perturbation with the largest eigenvalue is not always best at moderately large $\epsilon$. 

\begin{figure}
    \centering
\includegraphics[width=1\textwidth]{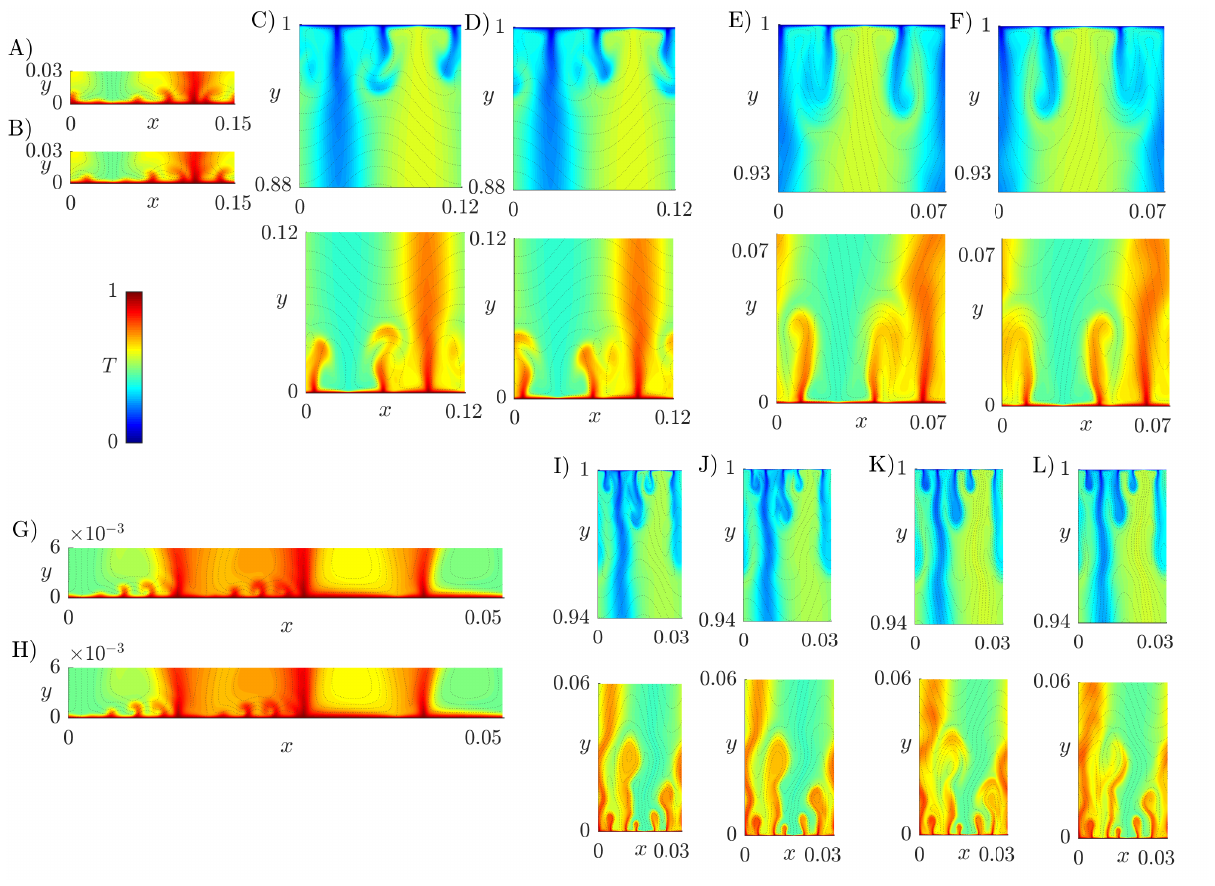}
    \caption{\footnotesize Snapshots of temperature fields corresponding to the flows with the largest Nu values at six (Pe, $\tau$Pe) combinations. Panels A--B, C--D, E--F, G--H, I--J, and K--L correspond to movies 3, 6, 11, 12, 15, and 16 respectively, and to the cases in figure \ref{fig:BestNuEpsModeFigure}A labeled with the same numbers. The black dotted lines show the instantaneous streamlines.} 
\label{fig:FlowTempMoviesFigure}
             \vspace{-.10in}
\end{figure}

To illustrate the effect of the flow perturbations at finite amplitude, in figure \ref{fig:FlowTempMoviesFigure} we show snapshots of the temperature fields for six of the cases in figure \ref{fig:BestNuEpsModeFigure}---those labeled 3, 6, 11, 12, 15, and 16, shown in panels A--B, C--D, E--F, G--H, I--J, and K--L respectively. Each pair of panels shows the temperature field at integer and half-integer multiples of the period $\tau$, i.e.~at opposite phases. The superposed black dotted lines show the instantaneous streamlines. The corresponding movies (with the same numbers as in figure \ref{fig:BestNuEpsModeFigure}) give a clearer sense of the motion, but the panels illustrate the basic features. 

Panels A and B show the temperature field for the top mode at Pe = 10$^5$ and $\tau$Pe = 10$^0$, which consists of two arrays of vortices moving along the bottom wall (see movie 3). Each array moves towards the hot plume that emanates on the right side near $x = 0.12$, one from the left and the other from the right (across the periodic boundary). The vortices create an array of small hot plumes along the bottom wall that grow and curl as they move toward the larger plume. The temperature field for the corresponding steady flow is shown in figure \ref{fig:StaticTempFig}G. Panels C and D of figure \ref{fig:FlowTempMoviesFigure} show the temperature field for the top mode at Pe = 10$^{5.5}$ and $\tau$Pe = 10$^{0.5}$, for which the steady flow is branched. The unsteady perturbation consists of chains of vortices that move around the eddies enclosed by the branches on the top and bottom walls (see movie 6). As a result, the three steady temperature plumes on the top and bottom walls in figure \ref{fig:StaticTempFig}H, two small and one large, oscillate from side to side in figure \ref{fig:FlowTempMoviesFigure}C--D. A similar side-to-side oscillation occurs in panels E and F, which correspond to the top mode at Pe = 10$^{6}$ and $\tau$Pe = 10$^{1}$. However, the flow mode is quite different from the previous one. Rather than a chain of vortices, we have one or two patches of vorticity that fill the entire horizontal width of the domain near the walls (see movie 11). These patches are zones of leftward and rightward shear flows along the walls.

The next set of panels, G and H, are for the top mode at the next higher Pe, 10$^{6.5}$, but a smaller $\tau$Pe, 10$^{-0.5}$. Here two chains of tiny vortices move along portions of the bottom wall, approaching two of the three larger hot plumes (shown for the steady case in figure \ref{fig:StaticTempFig}J) from the left. The resulting temperature fields resemble panels A and B. The last two pairs of panels, I--J and K--L, correspond to the largest Pe, 10$^{7}$, and two moderate $\tau$Pe, 10$^{0.5}$ and 10$^{1}$. In the steady case, figure \ref{fig:StaticTempFig}K, there are several plumes of different sizes on the top and bottom walls. The flow perturbations for I--J and K--L, shown in movies 15 and 16 respectively, are qualitatively similar. Vortices are predominantly at one wall (the top for I--J and the bottom for K--L), and alternate between small patches that move along the steady streamline branches, and larger regions of vorticity that cover the horizontal range near the wall. As previously, the most visible effect is the oscillation and in some cases pinch-off of the plumes. The net increase in Nu of 3-7\% relative to the steady case corresponds to an average thinning of the temperature boundary layers along the wall by the same amount. This relatively small thinning generally corresponds to the vortices moving colder fluid from the outer region towards the wall. 

\section{Conclusions \label{sec:Conclusions}}

We have calculated optimal unsteady flow perturbations for increasing the rate of heat transfer Nu from the values for the best known steady flows for the wall-to-wall problem, across a range of Pe (square root of flow power) from $10^2$ to $10^7$. The optimal perturbations are the leading eigenvectors of the Hessian matrix of Nu, the matrix of second derivatives with respect to amplitudes of perturbation modes. The resulting changes in Nu are given by the corresponding eigenvalues, which become positive starting at a Pe value between $10^{3.5}$ and $10^{4}$. In this regime, unsteady perturbations can outperform the steady flows.  

The same perturbation method was used previously in the case of optimal unidirectional flows through a heated channel. The eigenvalue distributions and eigenvector flow patterns have some similarities in the two cases. In both cases eigenvalues first become positive above a critical Pe near $10^3$--$10^4$. The number of positive eigenvalues increases from 1--2 to tens and hundreds as Pe increases to $10^5$--$10^7$. Positive eigenvalues occur in bands of flow periods $\tau$ that scale as Pe$^{-1}$, i.e. $\tau$Pe $\sim O(1)$. For the channel flows, the eigenvalues have a single maximum with respect to $\tau$Pe, at 10$^{0}$--10$^{0.5}$. For the wall-to-wall flows, there are two maxima, one at
$\tau$Pe = 10$^{0}$--10$^{0.5}$ and the other at $\tau$Pe = 10$^{1.5}$, which coalesce into a single maximum at Pe $\geq$ 10$^{5.5}$. 

For the channel flows, the corresponding eigenmodes are chains of vortices of alternating sign that move with the steady flow as traveling waves. At Pe $\leq 10^5$, where the steady wall-to-wall flows are rectangular convection rolls, and at $\tau$Pe $\leq$ 10$^{0.5}$, the wall-to-wall eigenmodes are also vortex chains near a wall, one per convection roll. At $\tau$Pe = 10$^{1}$, the wall-to-wall eigenmodes are instead pointlike vortices. At larger $\tau$Pe they become larger and more complicated vorticity distributions that spread vertically into the bulk flow.

At Pe $\geq 10^{5.5}$, where the steady wall-to-wall convection rolls undergo branching at the walls, the top eigenmodes are vortex chains along the walls at $\tau$Pe = 10$^{-0.5}$, and then along the eddies enclosed by the branches near the walls at larger $\tau$Pe. They again become more complex at $\tau$Pe $\geq$ 10$^{1}$, with a vorticity pattern that has some correlation with the branching pattern of the underlying steady flow.

At leading (quadratic) order, the time-averaged wall heat flux distributions show peaks near the vortex chains and other vorticity structures near the walls. 

We also solved for the temperature and Nu when the unsteady flow perturbations have moderate-to-large amplitudes. Unlike for the channel flows, a time-stepping method is inefficient for attaining the time-periodic solution to the unsteady advection-diffusion equation in the wall-to-wall case. Therefore we solved the equation in time-spectral form using the GMRES method with a block-Gauss-Seidel preconditioner. Convergence was rapid at smaller perturbation amplitudes ($\epsilon$) and smaller Pe, but often did not occur at larger values. Nonetheless, for large ranges of Pe and $\tau$Pe, we were able to find an interior maximum of Nu within the discrete set of $\epsilon$ used. The peak Nu ranged up to 7\% higher than that of the steady optimal flows. Due to computational time constraints, the search was limited to a somewhat sparse set of parameters. Given the sensitivities of the flows and temperature fields to small changes in parameters, it is possible that significantly larger increases in Nu could be found at nearby parameters.

\section*{Acknowledgments}
\vspace{-.25cm}
\noindent S.A. acknowledges support from the NSF-DMS Applied Mathematics program,
award number DMS-2204900.


\begin{thebibliography}{10}

\bibitem{layton1988history}
ET~Layton~Jr, JH~Lienhard, ET~Layton~Jr, and JH~Lienhard.
\newblock A history of the heat transfer division.
\newblock In {\em History of Heat Transfer: Essays in Honor of the 50th Anniversary of the ASME Heat Transfer Division, ET Layton, Jr., and JH Lienhard, eds., ASME, New York}, pages 1--23. ASME, 1988.

\bibitem{kaviany1994principles}
Massoud Kaviany.
\newblock {\em Principles of convective heat transfer}.
\newblock Springer, 1994.

\bibitem{webb2005enhanced}
Ralph~L Webb and NY~Kim.
\newblock {\em Enhanced heat transfer}.
\newblock Taylor and Francis, NY, 2005.

\bibitem{glezer2016enhanced}
Ari Glezer, Rajat Mittal, and Silas Alben.
\newblock Enhanced forced convection heat transfer using small scale vorticity concentrations effected by flow driven, aeroelastically vibrating reeds.
\newblock Technical report, Georgia Institute of Technology Atlanta United States, 2016.

\bibitem{yabe1996active}
Akira Yabe, Yasuo Mori, and Kunio Hijikata.
\newblock Active heat transfer enhancement by utilizing electric fields.
\newblock {\em Annual review of heat transfer}, 7, 1996.

\bibitem{laohalertdecha2007review}
Suriyan Laohalertdecha, Paisarn Naphon, and Somchai Wongwises.
\newblock A review of electrohydrodynamic enhancement of heat transfer.
\newblock {\em Renewable and Sustainable Energy Reviews}, 11(5):858--876, 2007.

\bibitem{fang2013active}
Ruixian Fang and Jamil~A Khan.
\newblock Active heat transfer enhancement in single-phase microchannels by using synthetic jets.
\newblock {\em Journal of Thermal Science and Engineering Applications}, 5(1):011006, 2013.

\bibitem{leal2013overview}
L~L{\'e}al, Marc Miscevic, Pascal Lavieille, M~Amokrane, Fran{\c{c}}ois Pigache, Fr{\'e}d{\'e}ric Topin, Bertrand Nogar{\`e}de, and L~Tadrist.
\newblock An overview of heat transfer enhancement methods and new perspectives: Focus on active methods using electroactive materials.
\newblock {\em International Journal of heat and mass transfer}, 61:505--524, 2013.

\bibitem{shank2023review}
Kyle Shank and Saeed Tiari.
\newblock A review on active heat transfer enhancement techniques within latent heat thermal energy storage systems.
\newblock {\em Energies}, 16(10):4165, 2023.

\bibitem{wang2025vortices}
Xiaojia Wang and Silas Alben.
\newblock How vortices enhance heat transfer from an oscillating plate.
\newblock {\em Journal of Fluid Mechanics}, 1013:A47, 2025.

\bibitem{hassanzadeh2014wall}
Pedram Hassanzadeh, Gregory~P Chini, and Charles~R Doering.
\newblock {Wall to wall optimal transport}.
\newblock {\em Journal of Fluid Mechanics}, 751:627--662, 2014.

\bibitem{sondak2015optimal}
David Sondak, Leslie~M Smith, and Fabian Waleffe.
\newblock {Optimal heat transport solutions for Rayleigh--B{\'e}nard convection}.
\newblock {\em Journal of Fluid Mechanics}, 784:565--595, 2015.

\bibitem{wen2020steady}
Baole Wen, David Goluskin, Matthew LeDuc, Gregory~P Chini, and Charles~R Doering.
\newblock Steady rayleigh--b{\'e}nard convection between stress-free boundaries.
\newblock {\em Journal of Fluid Mechanics}, 905:R4, 2020.

\bibitem{wen2022steady}
Baole Wen, David Goluskin, and Charles~R Doering.
\newblock Steady rayleigh--b{\'e}nard convection between no-slip boundaries.
\newblock {\em Journal of Fluid Mechanics}, 933:R4, 2022.

\bibitem{lohse2024ultimate}
Detlef Lohse and Olga Shishkina.
\newblock Ultimate rayleigh-b{\'e}nard turbulence.
\newblock {\em Reviews of modern physics}, 96(3):035001, 2024.

\bibitem{souza2016optimal}
Andre~N Souza.
\newblock {\em {An Optimal Control Approach to Bounding Transport Properties of Thermal Convection}}.
\newblock PhD thesis, University of Michigan, 2016.

\bibitem{souza2020wall}
Andre~N Souza, Ian Tobasco, and Charles~R Doering.
\newblock Wall-to-wall optimal transport in two dimensions.
\newblock {\em Journal of Fluid Mechanics}, 889:A34, 2020.

\bibitem{souza2015maximal}
Andre~N Souza and Charles~R Doering.
\newblock {Maximal transport in the Lorenz equations}.
\newblock {\em Physics Letters A}, 379(6):518--523, 2015.

\bibitem{souza2015transport}
Andre~N Souza and Charles~R Doering.
\newblock {Transport bounds for a truncated model of Rayleigh--B{\'e}nard convection}.
\newblock {\em Physica D: Nonlinear Phenomena}, 308:26--33, 2015.

\bibitem{tobasco2017optimal}
Ian Tobasco and Charles~R Doering.
\newblock Optimal wall-to-wall transport by incompressible flows.
\newblock {\em Physical review letters}, 118(26):264502, 2017.

\bibitem{doering2019optimal}
Charles~R Doering and Ian Tobasco.
\newblock On the optimal design of wall-to-wall heat transport.
\newblock {\em Communications on Pure and Applied Mathematics}, 72(11):2385--2448, 2019.

\bibitem{alben2023transition}
Silas Alben.
\newblock Transition to branching flows in optimal planar convection.
\newblock {\em Phys. Rev. Fluids}, 8:074502, Jul 2023.

\bibitem{motoki2018optimal}
Shingo Motoki, Genta Kawahara, and Masaki Shimizu.
\newblock Optimal heat transfer enhancement in plane couette flow.
\newblock {\em Journal of Fluid Mechanics}, 835:1157--1198, 2018.

\bibitem{motoki2018maximal}
Shingo Motoki, Genta Kawahara, and Masaki Shimizu.
\newblock Maximal heat transfer between two parallel plates.
\newblock {\em Journal of Fluid Mechanics}, 851:R4, 2018.

\bibitem{kumar2024three}
Anuj Kumar.
\newblock Three dimensional branching pipe flows for optimal scalar transport between walls.
\newblock {\em Nonlinearity}, 37(11):115011, 2024.

\bibitem{alben2024optimal}
Silas Alben.
\newblock Optimal wall shapes and flows for steady planar convection.
\newblock {\em Journal of Fluid Mechanics}, 984:A43, 2024.

\bibitem{alben2017improved}
Silas Alben.
\newblock Improved convection cooling in steady channel flows.
\newblock {\em Physical Review Fluids}, 2(10):104501, 2017.

\bibitem{alben_2017}
S.~Alben.
\newblock {Optimal convection cooling flows in general 2D  geometries}.
\newblock {\em Journal of Fluid Mechanics}, 814:484–509, 2017.

\bibitem{toppaladoddi2017roughness}
Srikanth Toppaladoddi, Sauro Succi, and John~S Wettlaufer.
\newblock Roughness as a route to the ultimate regime of thermal convection.
\newblock {\em Physical review letters}, 118(7):074503, 2017.

\bibitem{toppaladoddi2021thermal}
Srikanth Toppaladoddi, Andrew~J Wells, Charles~R Doering, and John~S Wettlaufer.
\newblock Thermal convection over fractal surfaces.
\newblock {\em Journal of Fluid Mechanics}, 907:A12, 2021.

\bibitem{marcotte2018optimal}
Florence Marcotte, Charles~R Doering, Jean-Luc Thiffeault, and William~R Young.
\newblock Optimal heat transfer and optimal exit times.
\newblock {\em SIAM Journal on Applied Mathematics}, 78(1):591--608, 2018.

\bibitem{song2023bounds}
Binglin Song, Giovanni Fantuzzi, and Ian Tobasco.
\newblock Bounds on heat transfer by incompressible flows between balanced sources and sinks.
\newblock {\em Physica D: Nonlinear Phenomena}, 444:133591, 2023.

\bibitem{alben2024enhancing}
Silas Alben, Shivani Prabala, and Mitchell Godek.
\newblock Enhancing heat transfer in a channel with unsteady flow perturbations.
\newblock {\em Physical Review Fluids}, 9(12):124503, 2024.

\bibitem{lamb1932hydrodynamics}
Horace Lamb.
\newblock {\em {Hydrodynamics}}.
\newblock Cambridge University Press, 1932.

\bibitem{hackbusch1981fast}
Wolfgang Hackbusch.
\newblock Fast numerical solution of time-periodic parabolic problems by a multigrid method.
\newblock {\em SIAM Journal on Scientific and Statistical Computing}, 2(2):198--206, 1981.

\bibitem{xiaolinwang2014thesis}
Xiaolin Wang.
\newblock {\em {A numerical study of vorticity-enhanced heat transfer}}.
\newblock PhD thesis, Georgia Institute of Technology, 2014.

\bibitem{de2025multigrid}
Hans De~Sterck, Stephanie Friedhoff, Oliver~A Krzysik, and Scott~P MacLachlan.
\newblock Multigrid reduction-in-time convergence for advection problems: A fourier analysis perspective.
\newblock {\em Numerical Linear Algebra with Applications}, 32(1):e2593, 2025.

\bibitem{boyd2001chebyshev}
John~P Boyd.
\newblock {\em Chebyshev and Fourier spectral methods}.
\newblock Courier Corporation, 2001.

\end{thebibliography}

\end{document}